\documentclass[10pt]{article}
\usepackage{amsfonts}
\usepackage{amsmath}
\usepackage{amssymb}
\usepackage{amsbsy}
\usepackage{deluxetable}
\usepackage[T1]{fontenc}
\usepackage[british]{babel}
\usepackage{latexsym}
\usepackage{mathtools}
\usepackage{setspace}
\usepackage{caption}
\usepackage{color}   
\usepackage{epsfig}
\usepackage{graphics}
\usepackage{graphicx}
\usepackage{xmpmulti}
\usepackage[varg]{txfonts}
\usepackage{aas_macros} 
\usepackage{astrobib} 
\def\i{\item}

\newcommand{\bed}{\begin{displaymath}}
\newcommand{\eed}{\end{displaymath}}
\newcommand{\bei}{\begin{itemize}}
\newcommand{\eei}{\end{itemize}}
\newcommand{\bef}{\begin{figure}}
\newcommand{\eef}{\end{figure}}
\newcommand{\ben}{\begin{enumerate}}
\newcommand{\een}{\end{enumerate}}
\newcommand{\beq}{\begin{equation}}
\newcommand{\eeq}{\end{equation}}
\newcommand{\ber}{\begin{eqnarray}}
\newcommand{\eer}{\end{eqnarray}}

\newcommand{\bb}{\bf B}

\newcommand{\edot}{\mbox{$\dot E$}}

\newcommand{\lsim}{\raisebox{-0.3ex}{\mbox{$\stackrel{<}{_\sim} \,$}}}

\newcounter{attnctr} 

\newcommand{\mathsym}[1]{{}}
\newcommand{\unicode}[1]{{}}
\begin{document}

\title{Neutron Star Physics in the SKA Era \\
       {\em An Indian Perspective}}
\author{Sushan Konar$^1$,
        Manjari Bagchi$^2$, 
        Debades Bandyopadhyay$^3$,\\ 
        Sarmistha Banik$^4$, 
        Dipankar Bhattacharya$^5$, 
        Sudip Bhattacharyya$^6$, \\
        R. T. Gangadhara$^7$,  
        A. Gopakumar$^6$, 
        Yashwant Gupta$^1$,  
        B. C. Joshi$^1$, \\
        Yogesh Maan$^8$, 
        Chandreyee Maitra$^9$, 
        Dipanjan Mukherjee$^{10}$, 
        Archana Pai$^{11}$, \\
        Biswajit Paul$^{12}$, 
        Alak K. Ray$^6$, 
        Firoza K. Sutaria$^7$ \\ 
        $^1$ NCRA-TIFR, Pune, India;
        $^2$ IMSc-HBNI, Chennai, India;
        $^3$ SINP, Kolkata, India;   \\
        $^4$ BITS-Pilani, Hyderabad, India; 
        $^5$ IUCAA, Pune, India;
        $^6$ TIFR, Mumbai, India; \\ 
        $^7$ IIAp, Bangalore, India; 
        $^9$ CEA, Saclay, France;  
        $^{10}$ RSAA-ANU, Canberra, Australia; \\
        $^8$ ASTRON, Netherlands Institute for Radio Astronomy, Dwingeloo, The Netherlands; \\
        $^{11}$ IISER, Thiruvananthapuram, India; 
        $^{12}$ RRI, Bangalore, India \\
{\em email : sushan@ncra.tifr.res.in}
}

\maketitle

\begin{abstract}
It is an  exceptionally opportune time for Astrophysics  when a number
of next-generation mega-instruments are poised to observe the universe
across  the entire  electromagnetic spectrum  with unprecedented  data
quality.  The Square  Kilometre Array (SKA) is undoubtedly  one of the
major components of this scenario.  In particular, the SKA is expected
to discover  tens of  thousands of  new neutron  stars giving  a major
fillip  to a  wide range  of  scientific investigations.  India has  a
sizeable  community  of scientists  working  on  different aspects  of
neutron star physics  with immediate access to both the  uGMRT (an SKA
pathfinder) and the recently launched X-ray observatory Astrosat.  The
current interests of the community largely centre around studies of -
a) {\em the generation of neutron stars and the SNe connection},
b) {\em the neutron star population and evolutionary pathways},
c) {\em the evolution of neutron stars in binaries and the magnetic fields},
d) {\em the neutron star equation of state},
e) {\em the radio pulsar emission mechanism}, and,
f) {\em the radio pulsars as probes of gravitational physics}.
Most of these studies are the main goals of the SKA first phase, which
is likely  to be  operational in  the next  four years.   This article
summarises the science  goals of the Indian neutron  star community in
the SKA era,  with significant focus on coordinated  efforts among the
SKA and other existing/upcoming instruments.
\end{abstract}

{\bf Keywords :}
    neutron stars : generation, population, eos, magnetic fields
  -- radio pulsar : emission -- radio pulsars : gravitational waves


\section{Introduction}

Since  the  first  detection  of  a  neutron  star  (NS)  as  a  radio
pulsar~\cite{hewis68}, we  now have some $\sim  2500$ objects detected
with  diverse  characteristic  properties  across  almost  the  entire
electromagnetic  spectrum~\cite{manch05a}.  A  consequence of  this is
the emergence  of a  large number  of distinct  observational classes.
Over the  years, the study of  this huge variety of  neutron stars has
mainly been  focused into three primary  areas - i) to  understand the
evolutionary   (or   otherwise)   connection  between   the   distinct
observational  classes,  ii)  to  understand  the  physical  processes
relevant in  and around  a neutron  star and iii)  to use  the neutron
stars as tools to understand certain aspects of fundamental physics.

All of these areas are expected to receive a tremendous boost with the
advent of new generation  instruments.  In particular, because neutron
star  astronomy depends  greatly on  radio observations,  the SKA  era
would  be of  paramount  importance.   Potential applications  include
pulsar search and survey, high precision pulsar timing to test gravity
theories  as  well  as  detect gravity  waves,  pulsar  magnetospheric
studies, characterisation  of transient  objects (like rotating radio
transients or RRATs),  and so
on.   These  will  provide  new  insights and  results  in  topics  of
fundamental importance.   The combination  of the high  spectral, time
and spatial  resolution, availability of large  number of simultaneous
beams in  the sky and  the unprecedented  sensitivity of the  SKA will
radically  advance  our  understanding  of  basic  physical  processes
operative in the vast population of  neutron stars and provide a solid
foundation for  the future  advancement of the  field.  The  extent of
such  impact  has recently  been  discussed  in  great detail  in  the
international                       neutron                       star
community~\cite{ska01,ska02,ska03,ska04,ska05,ska06,ska07,ska08,ska09}.
In this document  we focus on the areas of  particular interest to the
Indian scientists working  in different areas of  neutron star physics
in the SKA context.

\section{The Generation : NS - SNR associations}

Canonically neutron stars are created as end products of core collapse
supernovae  (CC-SNe).  To  date only  79 of  them have  been spatially
associated with  supernovae remnants (SNR).  While in most  cases, the
spin down  ages of the  pulsars match well with  the life time  of the
SNRs, there exist several objects of comparable age, for which no SNRs
have  been detected  in any  electromagnetic  band.  This  could be  a
selection  effect,  caused  by  the  sensitivity  limitations  of  the
(multi-waveband)  instruments,   or  it   could  be  due   to  factors
controlling the temporal  and spatial evolution and  emissivity of the
SNRs.   As the  supernova  shock propagates  through the  interstellar
medium (ISM), forming an SNR, its life time dominated by the following
four distinct phases.
\ben
   \i The  free expansion phase  - This  may last for  several hundred
   years and is dominated by thermal emission, wherein the mass of the
   swept up ISM dust  and gases is very small compared  to the mass of
   the ejecta ($M_{\rm ISM} \ll M_{\rm ejecta}$), with the temperature
   of  the fast  moving  medium being  in  the range  $T  \sim 10^6  -
   10^7$~K~\cite{cheva77}.
   \i The adiabatic (Sedov--Taylor) expansion  phase - This is reached
   when $M_{\rm  ISM} \simeq  M_{\rm ejecta}$, and  in this  phase the
   cooling ejecta also emits X-ray lines and thermal bremsstrahlung in
   the entire electromagnetic band from the radio to the X-rays.
   \i  The  ``snow-plow''  phase  - This  is  dominated  by  radiative
   cooling, and is reached when  $M_{\rm ISM} \gg M_{\rm ejecta}$, and
   $T \le 10^5$~K, leading to optical and UV line emission. This phase
   continues for $\sim 10^5$~year. 
   \i The final mixing  of the ISM and the ejecta  such that the shell
   breaks  up   in  clumps  (possibly  due   to  Rayleigh--Taylor  and
   Kelvin--Helmholtz instabilities)  and disperses  the SN-provender's
   material, and  the energy of  the shock  is dissipated in  the ISM.
%
\een
The duration for which the SNR remains in each phase, and its spectra,
are determined by the strength of SNe shock, the $M_{\rm ejecta}$, the
density of the ISM  (all of which in turn depend  on the properties of
the progenitor), and also on  the mass and thermodynamic properties of
the ISM.

In the  absence of  a central  compact object  (CCO), SNRs  are mostly
identified by  their morphology  (shell-like, plerionic,  or composite
structure)  and the  presence of  non-thermal (synchrotron)  emission,
emitted  mainly  by  highly   relativistic  electrons  accelerated  by
magnetic  fields  trapped in  the  remnant,  along with  some  diffuse
emission  from thermal  bremsstrahlung  processes.  While the  reverse
shock in a young  SNR serves as a good accelerator,  in older SNRs, the
radio  morphology (mainly  filamentary emission)  and luminosity  also
trace  out regions  of turbulence  induced amplification  of magnetic
fields   in   the  SNRs,   caused   by   interaction  with   the   ISM
clouds.  Further,  the  spectral  index of  the  synchrotron  emission
(especially  early  in  the  lifetime  of the  SNR),  can  be  changed
significantly by the  high energy cosmic rays (mainly  protons) in the
presence of strong  magnetic fields~\cite{jun99}.

There  exist several  radio pulsars  (PSR) with  spin down  ages below
$\tau \simeq 10^4$~year, with magnetic  field in the range of $10^{12}
< {\bf B}_{\rm surf} \simeq  10^{14}$~G (i.e.  from "ordinary" pulsars
to magnetars), which have no  observational evidence of any associated
SNRs (see table~[\ref{tab01}]).  Likewise, while the presence of a CCO
would confirm a CC-SNe, its absence in an SNR may also be because of a
high  kick velocity  imparted  to the  CCO at  birth,  or because  the
emission  is beamed  away from  us, with  any thermal  radiation being
below the detection threshold  of instruments in appropriate wavebands
(young CCOs emit thermally in the  0.2-10 keV band).  For example, the
nearest, recent  SNR, SN1987A,  shows no CCO  but Very Large Array (VLA)
observations do
show  some evidence  of  a  pulsar wind  nebular  (PWN). Finally,  the
paucity  of  SNR-PSR  associations  may  simply  be  because  (a)  the
canonical "age"  of SNRs is  being grossly over-estimated,  and/or (b)
the pulsar velocities at birth may have been severely under-estimated.
These raise  important questions about  galactic rate of all  types of
supernovae, the  poorly understood  physics of supernovae  and neutron
star (or magnetar) formation, and SNe shock-ISM interaction.  The high
spatial resolution of  SKA, and its higher sensitivity  should be able
to answer  at least some  of these, both  by probing the  structure of
Galactic  SNRs  down  to  higher  resolution  and  faintness,  and  by
discovering new SNRs  which may have been missed out  due to limits on
the current surveys.

The neutron stars associated with the CC-SNe fall into two categories;
the pulsar wind nebula and the central compact objects. In
young pulsars,  the rapid rotation  and large magnetic fields  (B $\ge
10^{12}$  G)   accelerate  particles  and  produce   energetic  winds,
resulting in spin-down of the  neutron star. Confinement of the pulsar
winds in the surrounding medium generates luminous PWN seen all across
the electromagnetic  spectrum, shining  most prominently in  radio and
X-rays.  First proposed  by \citeN{rees74}, morphology of  the PWN can
provide  crucial information  on the  properties of  the outflow,  the
interacting ambient medium and the  geometry of the pulsar powering it
(see  \citeNP{gaens06},  and  references  therein).  In  the  innermost
regions,  relativistic outflows  in the  form  of torus  and jets  are
formed; the  geometry of which  reveals the orientation of  the pulsar
spin  axes and  can  provide  information on  the  formation of  kicks
imparted in  the moments following their  formation.  The larger-scale
structures of the  PWN provide insights on the  ambient magnetic field
and signatures of interaction with the supernova ejecta. The evolution
of a PWN inside an SNR follows three important evolutionary phases: an
initial free-expansion in the  supernova ejecta, the collision between
the PWN and SNR reverse shock,  which crushes the PWN subjecting it to
various instabilities, and eventually subsonic re-expansion of the PWN
in the shock heated  ejecta~\cite{gelfa07} and references therein). An
additional later  phase might  also be  identified, when  the pulsar's
motion becomes supersonic at it  approaches the shell of the supernova
remnant and it acquires a bow-shock morphology~\cite{swalu04}.

CCOs are  X-ray point-like  sources found at  the centre  of supernova
remnants~\cite{pavlo05}.  They have  a purely  thermal X-ray  spectrum
with KT  $\sim$ 0.2--0.5  keV and luminosity  $\sim$ $10^{33}-10^{34}$
erg s$^{-1}$.  No radio/Gamma-ray  counterparts have been found. There
is also no evidence of  an extended nebular emission surrounding these
objects.

The higher sensitivity  and spatial resolution of the  SKA would allow
simultaneous imaging of SNR/PWN, along-with time series observations of
any plausible  associated pulsar candidate. For  example, the primary
beam at Band-2 of the proposed  SKA-mid is $\sim 50'$, which is larger
than the apparent size of most of the known SNRs~\cite{gre14}.  At the
same time,  with the longest  baseline of  about 160 km,  high dynamic
range images (which are not confusion limited) of these sources can be
obtained with a resolution better than a third of an arc-second.  Deep,
high resolution images  obtained with the SKA are  therefore likely to
(a) discover faint emission from the  SNRs and (b) increase the number
of pulsar-PWN  associations. Coupled  with simultaneous  sensitive time
resolution  searches  with  the  SKA,  these can  be  very  useful  in
constraining       evolution      and       dynamics      of       such
objects~\cite{cgd+11,csd+13,scj+15}.

\section{The Population}

\subsection{Classification}

Even  though  some  2500  neutron stars  with  diverse  characteristic
properties have been observed, interestingly the processes responsible
for the  generation of  the observed emitted  energy are  basically of
three types.   This leads  to a simple  classification of  the neutron
stars   according   to   the    nature   of   energy   generation   in
them~\cite{konar13}.   Primarily,  we  have  -  the  rotation  powered
pulsars (RPP) powered by the loss of rotational energy due to magnetic
braking;  and  the  accretion  powered pulsars  (APP)  where  material
accretion from  a companion gives  rise to energetic  radiation.  Then
there is the new class of internal energy powered neutron stars (IENS)
(for want  of a better name)  where the emission is  suspected to come
from their internal  reservoir of energy (be it that  of a very strong
magnetic   field   or  the   residual   heat   of  a   young   neutron
star)~\cite{kaspi10}.

%
\bef
   \begin{center}
      \includegraphics[angle=-90,width=15.0cm]{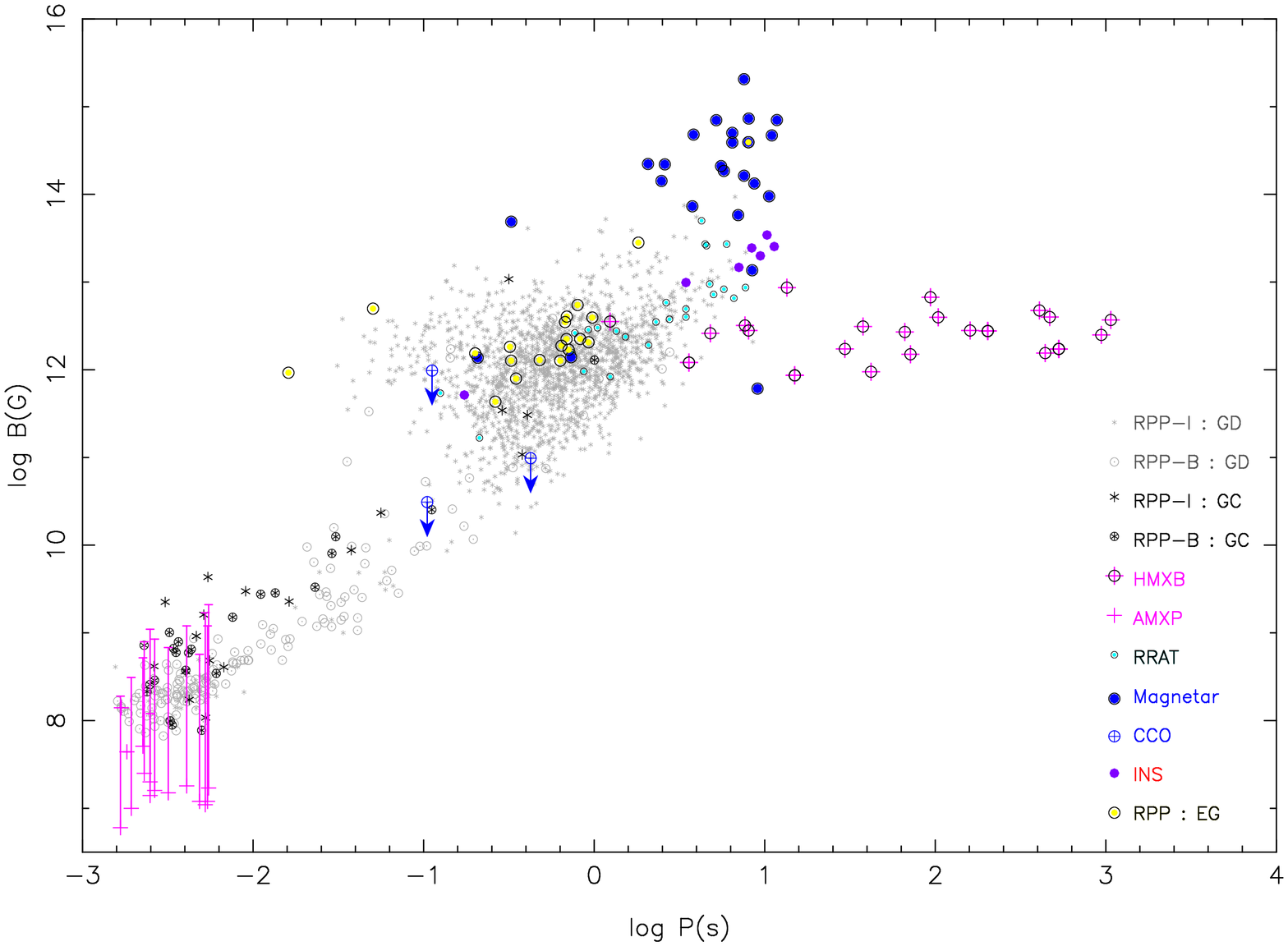}
   \end{center}
\caption{The Menagerie : Different observational  classes of neutron stars
shown in the spin-period vs. surface magnetic field ($P_s-B_s$) plane.  
{\bf \em Legends :} I/B - isolated/binary, GC - globular cluster, 
GD - galactic disc, EG - extra-galactic objects \\
{\bf \em Data :} RPP - ATNF pulsar catalogue,
RRAT - {\tt http://astro.phys.wvu.edu/rratalog/},
Magnetar - {\tt http://www.physics.mcgill.ca/~pulsar/magnetar/main.html},
AMXP - Patruno \& Watts (2012), Mukherjee et al. (2015);
HMXB - Caballero \& Wilms (2012),
INS - Haberl (2007), Kaplan \& van Kerkwicjk (2009);
CCO - Halpern \& Gotthelf (2010), Ho (2013).
}
\label{fig01}
\eef
\nocite{patru12c,mukhe15,cabal12,haber07,kapla09a,halpe10,ho13}

\ben
   \i {\bf  Rotation Powered Pulsars (RPP)  :} - As of  now, there are
   about three types that fall into these category.
   \ben
      \i These are mainly the classical  radio pulsars ({\bf PSR} - $P
      \sim 1$~s, $B \sim 10^{11}  - 10^{13.5}$~G), powered by the loss
      of rotational energy due to magnetic braking.
      \i  Among  the  current  sample of  2000+  radio  pulsars  known
      primarily in our Galaxy, millisecond radio pulsars ({\bf MSRP} -
      $P  \lsim 20$~ms,  $B \lsim  10^{10}$~G) now  number almost  200
      \cite{lorim09}.  This famous (and  initially the only) sub-class
      of  RPPs,  have   different  evolutionary  histories,  involving
      long-lived binary  systems and  a `recycling'  accretion episode
      reducing    both    the    spin-period    and    the    magnetic
      field~\cite{tauri11}.
      \i The mildly  recycled pulsars ({\bf MRP}),  defined as objects
      with $P \sim  20 - 100$~ms and $B_p <  10^{11}$~G.  The rotating
      radio transients ({\bf RRAT}) are  also likely to be a sub-class
      of RPPs;  suspected to be extreme  cases of nulling/intermittent
      pulsars.
   \een
   \i {\bf Accretion Powered Pulsars (APP)  :} - Depending on the mass
   of the donor star these  are classified as High-Mass X-ray Binaries
   (HMXB) or Low-Mass X-Ray Binaries (LMXB).
   \ben
      \i Neutron  stars in HMXBs  typically have $B_p  \sim 10^{12}$~G
      and $O$  or $B$ type companions  and mostly show up  as an X-ray
      pulsars  ({\bf  XRP})~\cite{cabal12}.They   have  a  hard  X-ray
      spectrum  (KT  $>$  15  keV).  The  spectrum  exhibits  possible
      signatures  of interaction  of  the accreted  material with  the
      strong  magnetic  field of  the  neutron  star  in the  form  of
      Comptonisation, and the presence of broad absorption lines known
      as  Cyclotron Resonance  Scattering Features  (CRSF). CRSFs  are
      found  in more  than 20  XRPs,  and is  a direct  tracer of  the
      magnetic field  strength of  the neutron  star from  the 12-B-12
      rule:   $E_{c}$  =   11.6  keV   $\times  \frac{1}{1+Z}   \times
      \frac{B}{10^{12}}$ G;  $E_{c}$ being the centroid  energy, z the
      gravitational red-shift and B the magnetic field strength of the
      neutron star.
      \i LMXBs, on the other hand, harbour neutron stars with magnetic
      fields significantly  weakened ($B \lsim 10^{11}$~G)  through an
      extended phase  of accretion.  Physical process  taking place in
      such accreting  systems are manifested as  - thermonuclear X-ray
      bursts;    accretion-powered   millisecond-period    pulsations;
      kilohertz quasi-periodic  oscillations; broad  relativistic iron
      lines;  and  quiescent  emissions~\cite{bhatts10}.   These  have
      given rise to two exciting observational classes in recent times
      - the accreting  millisecond X-ray pulsars ({\bf  AMXP}) and the
      accreting millisecond X-ray bursters ({\bf AMXB}).
   \een
   \i {\bf Internal  Energy Powered Neutron Stars :}  - The connecting
   link between these classes is the fact that the mechanism of energy
   generation is  not obvious  for any  of them. For  most part  it is
   suspected that  the decay  of a strong  magnetic field  or residual
   cooling might be responsible for the observed emission.
   \ben
       \i {\bf  Magnetars} are thought  to be young,  isolated neutron
       stars and they shine because of the decay of their super-strong
       magnetic  fields and  are actually  related to  soft gamma  ray
       repeaters ({\bf SGR}) and  anomalous X-ray pulsars ({\bf AXP}).
       It is believed that the main  energy source of these objects is
       the   decay   of   super-strong   magnetic   fields   (magnetar
       model)~\cite{thomp96}.
       \i The  handful of X-ray  bright compact central  objects ({\bf
         CCO}) are characterised by absence both of associated nebulae
       and of counterparts at  other wavelengths and exceptionally low
       magnetic fields  ($B \sim  10^{10}$~G).  It has  been suggested
       that a regime of  hypercritical accretion immediately after the
       birth  of the  neutron star  could bury  the original  field to
       deeper regions of the crust~\cite{vigan12}.
       \i  The seven  isolated  neutron stars  ({\bf INS}),  popularly
       known  as {\em  Magnificent Seven},  are optically  faint, have
       blackbody-like  X-ray  spectra  ($T \sim  10^6$~K),  relatively
       nearby and have long spin-periods  ($P \sim - 10$~s).  They are
       probably  like ordinary  pulsars  but a  combination of  strong
       magnetic  field  and spatial  proximity  make  them visible  in
       X-rays.
   \een
\een

One of the  prime challenges of neutron star research  has always been
to find  a unifying theme  to explain  the menagerie presented  by the
disparate  observational classes  (shown in  Fig.\ref{fig01}]) of  the
  neutron stars.  The  magnetic field, ranging from  $10^8$~G in MSRPs
  to  $10^{15}$~G in  magnetars, has  been central  to this  theme. It
  plays an  important role in  determining the evolution of  the spin,
  the radiative properties and the interaction of a neutrons star with
  its surrounding  medium.  Consequently, it  is the evolution  of the
  magnetic field which link these classes.

Some of  the evolutionary  pathways have  now become  well established
through  decades of  investigation by  a number  of researchers.   For
example, the connection between the  ordinary radio pulsars with their
millisecond  cousins  via  binary  processing is  now  an  established
evolutionary  pathway~\cite{bhatt02,konar13,konar16c}.   On the  other
hand, to understand the connection between different types of isolated
neutron stars a detailed theory  of magneto-thermal evolution is being
developed   only  in   the   last  few   years  (see   \citeN{pons09};
\citeN{kaspi10}; \citeN{vigan13} for details  of these model and other
references).    Therefore,  it   appears  that   a  scheme   of  grand
unification,  encompassing all  the varieties,  has started  to emerge
now. And  it is expected  that in the SKA  era important gaps  in this
unification scheme could be filled.

\subsection{Radio Pulsars : Statistical Studies}

Timing irregularities seen  in radio pulsar rotation rates  are of two
kinds - a) timing noise  : continuous, noise-like fluctuations; and b)
glitches.  The abrupt  cessation of pulsed radio  emission for several
pulse periods, observed  in some hundred odd pulsars, is  known as the
phenomenon of nulling.   The nature and degree of  this nulling varies
from one  pulsar to  the other. We  undertake a  comprehensive 
statistical study of these pulsars and 
also include the intermittent  pulsars and the
RRATs in  our study. Recently,  it has  been suggested that  there may
exist  a  trend for  nulling  activity,  going from  ordinary  nulling
pulsars to intermittent pulsars  to RRATs. Here we try to quantify the nulling
behaviour to check for any  difference between these different classes
of pulsars. With that  aim we find the proximity of  a given object to
the  death-line.   We  find  that   for  any  assumed  death-line  the
statistical distribution of  q d for ordinary nulling  pulsars is very
different from that of the RRATs.  A number of scenarios were proposed
to explain the new observational class of RRATs. Assuming them to be a
completely  separate   population  of  neutron  stars   leads  to  the
conclusion that  in such a  situation the inferred number  of Galactic
neutron stars would be totally at variance with the Galactic supernova
rates~\cite{keane10b}.  They have also been  conjectured to be part of
the  normal  pulsar  intermittency  spectrum -  the  case  of  extreme
nulling,   which  have   been   shown  not   to   hold  much   promise
recently~\cite{konar16d}.

A glitch is a timing irregularity of radio pulsars, marked by a sudden
increase in the  spin-frequency $\nu$ which may  sometimes be followed
by a  relaxation towards  its unperturbed  value.  These  are probably
caused by  sudden and  irregular transfer of  angular momentum  to the
solid crust  of the neutron  star by a super-fluid  component rotating
faster;  or by  the  crust  quakes. It  has  been  suggested that  the
bimodality seen within the range  of glitch values actually pertain to
these two  separate mechanisms~\cite{yu13}.  Our  statistical analysis
supports  the  conjecture  that  there  exist  more  than  one  glitch
mechanism corresponding  to different intrinsic energies.   We suggest
that  mechanisms responsible  for glitches  are perhaps  different for
different  energy regimes,  originating  in different  regions of  the
star~\cite{konar14}.

A detailed discussion about the statistical studies can be found in
Arjunwadkar, Kashikar \& Bagchi (this volume).

\subsection{Radio Pulsars : Synthesis Studies}

Population synthesis study is the  effort to understand the cumulative
properties of neutron  stars in different `classes' and  to reveal the
underlying physics behind such  observed properties. One simple method
of population synthesis study is the so called `snapshot' method where
people  study the  observed properties  of  one or  more `classes'  or
`sub-classes' \cite{hui10,konar10,papit14}.  This method is not always
sufficient to reveal the true  characteristics of a `class' of neutron
stars, mainly because  of our limitations in  observing neutron stars.
So   the  detailed   `full  dynamical'   method  becomes   unavoidable
\cite{bhatt92,fauch06,story07,ridle10}.   In  this method,  one  first
chooses a set  of initial parameters for the objects  belonging to the
`class'  or `classes'  under investigation  (Monte-Carlo simulations),
then model  the evolution  of the objects  with chosen  parameters, as
well  as  their  observability,  i.e. the  probability  of  detection.
Finally one  justifies the choice of  initial parameters, evolutionary
models,  etc.,  by  comparing   the  `observable'  properties  of  the
synthetic  set  of  objects  with  the  observed  properties  of  such
objects. This  method sometime  reveals interesting properties  of the
objects under investigation, e.g.,  the study by \citeN{bhatt92} first
established  the fact  that the  magnetic field  of isolated  rotation
powered radio pulsars does  not decay. Sometimes, methods intermediate
between the `snapshot' approach and the `full dynamical' approach have
been used \cite{lorim93,hanse97,bagch11,gullo14}.   Note that, because
of the complexity  of the `full dynamical'  approach and uncertainties
in initial conditions,  this method is more popular  to study isolated
radio pulsars with  the initial point of the evolution  started at the
birth of  the neutron stars. In  principle, one can perform  the study
starting from  the zero age main  sequence phase of the  progenitor of
the neutron  star, as  theory of  evolution of  massive stars  is well
known. Similarly, there are many studies on evolution and formation of
neutron                stars                in                binaries
\cite{bhatt91a,verbu93,porte98,dewi06,belcz08,kiel08,tauri13},   where
the evolution and  detectability of binary radio pulsars  had not been
studied.  Thus population  synthesis remains an open  avenue, which is
likely to shed  more light on the physics of  different sub-classes of
rotation powered pulsars (for example, the  RRATs) which are yet to be
understood  properly. Population  synthesis will  be useful  for other
areas  of research  as well,  including gravitational  wave astronomy,
short  GRBs  etc.   Moreover,  there  is  enough  scope  of  employing
population  synthesis methods  to  other `classes'  of neutron  stars,
e.g., APPs.

It  is  expected  that  the  SKA will  lead  to  discoveries  of  more
RPPs~\cite{smits09,smits11}. This  larger data  set will enable  us to
test  existing and  future  population synthesis  studies  as well  as
explore  evolutionary  pathways.   While  phase-1 of  the  SKA-low  is
proposed to  have 500 simultaneous  beams, the SKA-mid will  have 1600
beams within its large primary beam.  The full SKA is proposed to have
even  larger number  of  beams.  This  makes the  SKA  a rapid  survey
instrument. With up-to  1800~s {\em dwell time}  (length of integration
per position) and  96~MHz instantaneous bandwidth at  SKA1-low and 300
MHz instantaneous  bandwidth at  SKA1-mid, the first  phase of  SKA is
expected to find about 10000 new pulsars with 1500 MSRPs. The full SKA
is     expected     to     triple      the     number     of     these
discoveries~\cite{ska03}. These discoveries are  likely to uncover new
classes of  neutron stars  including pulsars  with stellar  mass black
hole (SMBH) companions, pulsars within a parsec of Galactic Centre and
probably even some  unexpected/unknown sub-classes.  The unprecedented
increase  in the  sample  is likely  to result  in  finding more  {\em
  missing link} objects to  firmly establish the evolutionary pathways
between these classes.

\bef
\begin{center}
\includegraphics[width=0.75\columnwidth]{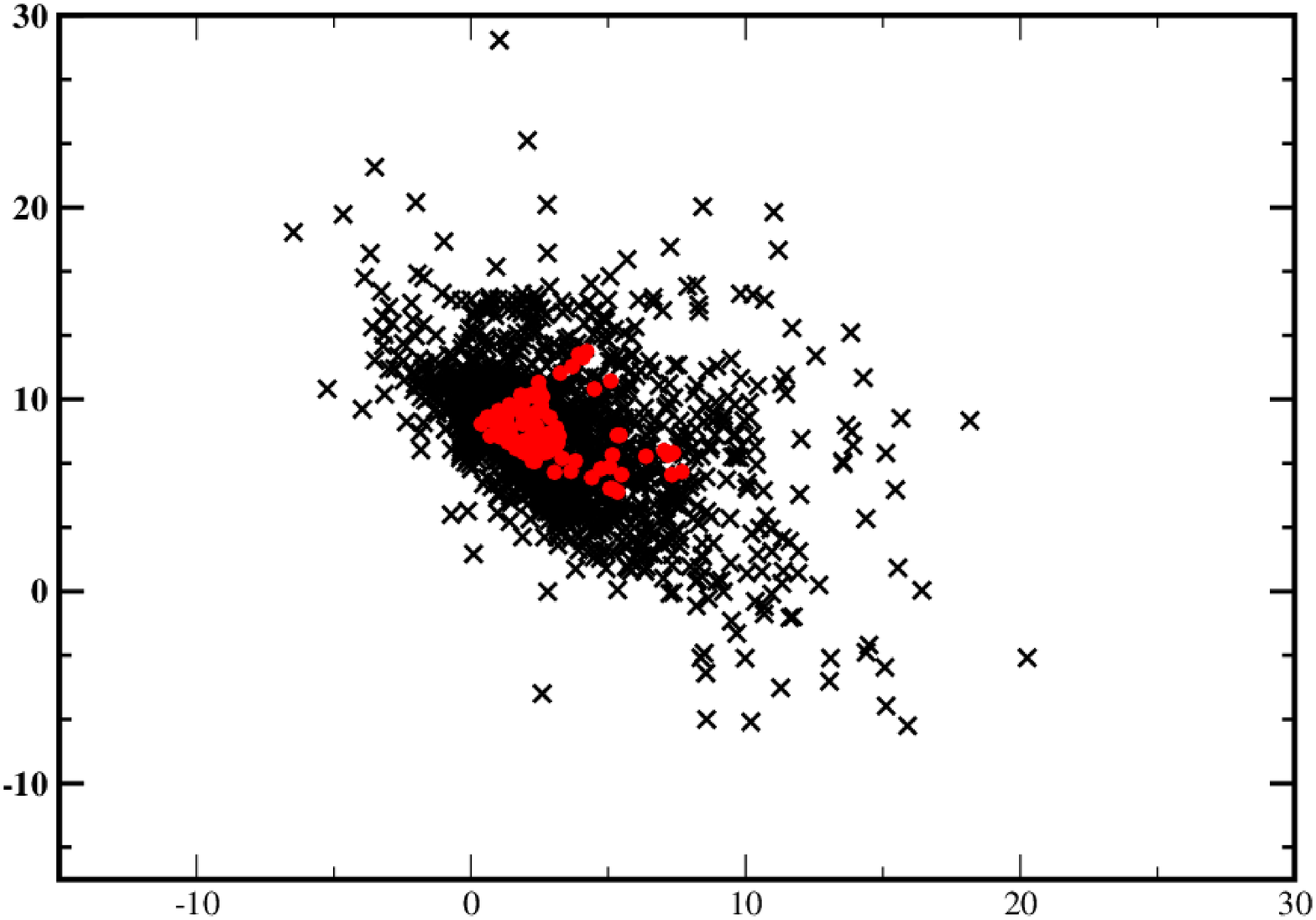}
\caption{Distribution of  expected RPP discoveries projected  onto the
  Galactic plane.  The Galactic centre is  at the origin and  the axes
  logarithmic distance in units of  Kpc.  Ordinary radio pulsars (PSR)
  are marked  with a  black $x$, while  the millisecond  radio pulsars
  (MSRP) are marked with filled red  circles.  This is from a proposed
  survey  of  the  north  Galactic  plane  obtained  using  population
  synthesis methods.  The proposed survey will use an incoherent array
  of  25   uGMRT  antennae  with   the  new  300-500   MHz  receivers,
  commissioned in the recent upgrade.}
\label{psrsrch}
\end{center}
\eef

The      GMRT     has      been      widely      used     for      RPP
surveys~\cite{fgr+04,gmg+05,jml+09}  at low  frequencies.  The  recent
upgrade provides an opportunity to use 200~MHz instantaneous bandwidth
between 300-500~MHz  for a rapid survey  of the sky, using  the GMRT's
$\sim  80'$ primary  beam.   The  frequency range  for  such a  survey
overlaps that  of the  SKA1-low and  Band-1 of  the SKA1-mid,  and can
provide useful inputs  for the eventual SKA1 pulsar  surveys with both
the instruments. At  present using an incoherent array  of 25 antennas
of the GMRT, with 4096 channels  across the 300$-$500 MHz bandwidth, a
sampling time of about 80~$\mu$s and a dwell time of 300~s, the entire
northern sky can  be covered in about 1100 hours  with the uGMRT. Even
with a  modest 200-250  hours of  observing time per  year, it  can be
completed well in advance of  the commencement of science observations
for the phase~1 of SKA. This  survey is expected to discover something
like 800 new normal pulsars and  20 new MSRPs according to simulations
from  population  synthesis  models~\cite{blr+14}.  Evidently  such  a
study, besides providing immediate science  returns for the uGMRT, has
the potential for becoming a pathfinder survey for pulsar science with
the  SKA, commensurate  with the  {\em ``pathfinder''}  status of  the
uGMRT.   The expected  distribution  of pulsars  from  this survey  is
indicated in Fig.ref{psrsrch}.

A  survey with  much higher  sensitivity  can be  performed using  the
fourteen antennas in the inner central  square of the GMRT in a phased
array. This would have a potential of discovering 2000 ordinary and 90
millisecond  pulsars  with 180~s  dwell  time.   Such a  survey  would
require   half   the  observing   time   with   much  larger   science
returns. However, for a greater survey speed, several beams in the sky
are needed and a new multi-beam receiver needs to be developed for this
purpose. Either way these proposed surveys will allow for devising new
observing and  search strategies.  Moreover, properly  designed, these
surveys would  help train young students  thus being useful both  as a
manpower  development effort  for the  SKA  as well  as preparing  the
community  for fruitful  exploitation  of data  from  the SKA  pulsar
surveys in future.

\subsection{Black-widows \& Red-backs : super-Jupiter companions}

Observation of  molecular lines in  the ablated wind  of super-Jupiter
companions  of  black-widow and/or  red-back  pulsars  can provide  an
interesting line of  investigation into the physics  of these systems.
The black-widow and  the red-back pulsars are binary MSRPs  but are so
named because they  are in the process of  destroying their companions
through strong pulsar winds.  The first black-widow pulsar observed is
PSR  B1957+20~\cite{fruch88,fruch90},   a  millisecond   radio  pulsar
($P_{spin} = 6.2 \;\rm ms$) ablating  its companion in a binary system
($P_{orb}     =    9.17     \;    \rm     hr$).     It     has    been
suggested~\cite{kluzn88,phinn88}  that  strong  gamma-ray  irradiation
from the  millisecond pulsar drives  the wind from the  companion star
and gives  rise to a  bow shock between  the wind and  pulsar magnetic
field. The  pulsar is  likely to  be left  as an  isolated millisecond
pulsar after few  times $10^8 \; \rm yr$.  The  observability of these
binary pulsars in  the black widow state implies that  the lifetime of
this transitory phase cannot be much shorter.  Given the strong pulsar
radiation and the relativistic  electron-positron outflow ablating its
companion to  drive a comet-tail like  wind, it is feasible  to search
for absorption lines in radio spectra.

Detection of  OH line  absorption through  radio line  spectroscopy in
intra-binary space around a few  millisecond pulsars has recently been
suggested~\cite{ray15}.    A   detection   may  lead   to   a   better
understanding of past evolutionary history  of such systems as well as
the composition of the companions  themselves.  The knowledge may help
us  to discern  formation scenarios  of ultra  low mass  companions of
pulsars.   Detection of  OH lines  would  lead to  constraints on  the
present state,  e.g.  the  mass and  radius of  the companion.   OH is
usually formed by  the dissociation of $\rm H_2 O$  molecule, so it is
often considered a proxy for water,  the most sought after molecule in
the  exoplanet context.   Black  widow and  red-back  pulsars have  the
characteristics  that make  them good  targets for  OH line  detection
observations since pulsar timing and  optical observations allow us to
determine  their geometry,  including the  eclipsing region,  well and
their short orbital periods allow  for repeated observations over many
orbits within reasonable time-lines to build up gated exposure time on
the radio pulses near eclipse ingress  and egress.  The geometry for a
pulsar beam passing through the ablated wind of a companion orbiting a
typical black widow pulsar is shown in Fig.\ref{fig:eclipse-BWPSR}.
\bef
\begin{center}
\includegraphics[width=0.5\columnwidth]{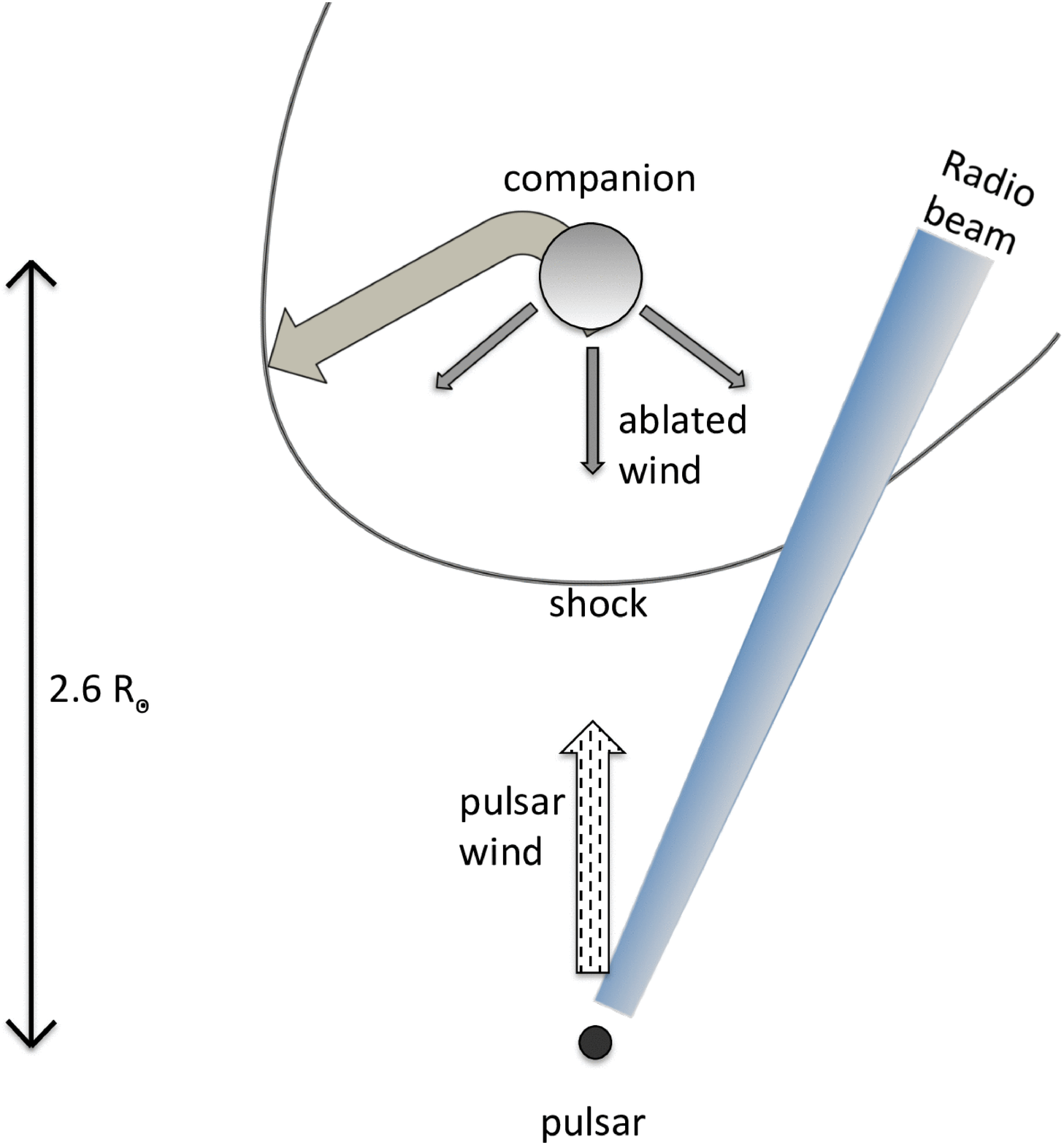}
\caption{The geometry  of a  pulsar beam  passing through  the ablated
  wind from the super-Jupiter companion orbiting a black widow pulsar.
  The wind may be rich in Oxygen  if the companion is the remnant of a
  star that has generated Carbon or Oxygen during its own evolutionary
  process.}
\label{fig:eclipse-BWPSR}
\end{center}
\eef

The intended targets  would have to include  {\it binary, millisecond}
pulsars with  high spin-down power  and moderate radio  flux densities,
with ultra-low mass (or even comparable to Jupiter mass) companions in
close orbits  ($P_{orb} \sim 2- 14  \; \rm hr$).  Their  moderate mean
flux densities and their small orbital dimensions would make them good
targets for OH  absorption studies against their pulsed  flux.  It has
been argued that  even with existing telescopes like  GBT, Arecibo and
Parkes it  should be possible to  detect the OH line  from black-widow
and   red-back   pulsars~\cite{ray15}.    Such  lines   in   absorption
spectroscopy or maser emission in  the (ISM) has already been detected
for a few pulsars~\cite{stani03,weisb05,minte08}.

Future  telescopes like  the Square  Kilometre Array  (SKA) will  have
typically ten (SKA1-Mid) to hundred  times (full SKA) flux sensitivity
compared to GBT.   With SKA-1 increased sensitivity  in L-band (Band 2)
using low  noise amplifiers,  it will  be possible  to carry  out such
molecular line detections down to a  pulsar mean flux density level of
$0.7\;  \rm mJy$  or  lower,  opening up  several  other known  pulsar
targets  for  similar studies.   In  addition,  many newly  discovered
pulsar targets will become available for studies of composition of the
winds from the companion.

\section{The Evolution}

\subsection{In HMXBs}

While  the double  neutron  star  (DNS) binaries  and  the  yet to  be
discovered NS-BH  binaries are the  most promising candidates  for the
gravitational wave  detectors, formation rates of  such systems depend
crucially on  the formation, co-evolution,  and survival rates  of the
massive binary stellar systems.  One important phase of such binaries,
in which  their evolution can  be measured  with high accuracy  is the
HMXB phase in which one component  has evolved to a compact star, most
often a high  magnetic field neutron star. The  accreting neutron star
in such a system is an  X-ray pulsar that enables accurate measurement
of the  orbital parameters and  its evolution.   In almost all  of the
accreting X-ray pulsars with  super-giant companion stars, the orbital
evolution time scale  is found to be short, less  than a million year;
suggesting a tidal force  driven orbital evolution~\cite{paul11}.  The
evolution of  such systems, leading to  the formation of DNS  or NS-BH
binaries  in some  of them  therefore  must account  for the  measured
orbital decay  rates of such systems.   The orbital decay may  lead to
more  compact  configuration which  will  increase  the survival  rate
during the second  supernova explosion and the decay may  also lead to
complete spiral-in which  will lead to a single  remnant.  The orbital
evolution of only a handful of such systems have been followed so far.
Measurements of orbital decay of a  large number of HMXBs are expected
to  be  performed  with  Astrosat   and  other  planned  X-ray  timing
instruments.  

To    date    only    a     single    double    pulsar    system    is
known~\cite{bdp+03,lbk+04}.   Twelve  double   neutron  star  binaries
(DNS), including the double pulsar  system, form a rather small sample
of   such  systems~\cite{bp16}.    These  allow   for  accurate   mass
measurements and  potentially radius measurement as  well, making them
important  candidates  for constraining  the  equations  of state,  as
discussed later  (See Section \ref{eosns}). X-ray  measurements of the
HMXBs would  have an impact on  estimating the rates of  DNS and NS-BH
binaries. This would affect searches  with the SKA and eventually have
an impact on  the event rate estimate of  gravitational wave detectors
as well.

\subsection{In LMXBs, Transition Pulsars}

MSRPs  are thought  to  be spun  up  by a  billion-year-long phase  of
angular momentum transfer  in LMXBs~\cite{alpar82,radha82}. While this
hypothesis  is widely  accepted,  till recently,  a direct  connection
between MSRPs and LMXBs  was not established, because radio pulsations
were not observed  from an LMXB. This is  very important to understand
the binary  evolution and accretion  processes, because the  extent of
spinning  up  depends on  (1)  the  evolution  of accretion  rate  and
structure due to  the evolution of the binary  properties, (2) whether
accretion primarily  happens via  a geometrically thin  accretion disk
and  hence transfers  a large  amount  of angular  momentum, (3)  what
fraction of  the accreted matter leaves  the system via  jet and wind,
(4) whether  accretion happens continuously or  intermittently, and so
on. Observation of radio pulsation  from an X-ray binary also gives us
a unique  opportunity to  study how pulsar  radiation and  pulsar wind
nebula  affect the  accretion process.   Recently, three  sources (PSR
J1023+0038,  PSR J1227-4853, PSR  J1824-2452I) have  shown transitions
between     the     radio     pulsar     phase    and     the     LMXB
phase~\cite{archi09,roy15,papit13}.   Each of the  three systems  is a
Red-back system, that is a binary stellar system with an ms pulsar and
a  main-sequence star  rotating around  each other.  These discoveries
have confirmed that such sources can change between radio pulsar state
and LMXB state back  and forth.  Moreover, various low-intensity X-ray
states have  recently been reported  for these sources~\cite{linar14}.
So, as  mentioned earlier, the  systems showing both radio  pulsar and
LMXB   phases  can  be   very  useful   to  probe   binary  evolution,
accretion-ejection mechanism, and so on.  In order to achieve this, it
is essential to  discover as many such systems  as possible, and study
them in  both radio and X-rays  in various intensity  states.  So far,
roughly 18 red-back  systems are known. SKA will  increase this number
by at least  a factor of a few~\cite{ska03}.  Moreover, the two phases
may be discovered even  from non-red-back sources.  The radio pulsation
phase of  this increased population  should be observed  and monitored
with  SKA. When  the radio  pulsation disappears,  or when  the source
X-ray  intensity  increases as  detected  with  X-ray monitors,  X-ray
pointed  observations should be  done.  This  way the  above mentioned
scientific problems can be  addressed very effectively by observations
with SKA and proposed X-ray observatories such as Astrosat and Athena.

\subsection{Evolution of the magnetic fields}

As mentioned in section 3.1, the evolution of magnetic field is one of
the central ingredients in  understanding the interconnections between
different observational classes of the neutron stars. The evolution of
the magnetic  field in accreting  neutron stars has been  of sustained
interest  to  a  number  of  Indian scientists  over  the  years.  The
following summarises the recent efforts in this area which is expected
to get a fillip with the  expected increase in the number of transient
pulsars  (like   black-widow  \&  red-back,  for   example)  and  other
categories in the SKA era.

In accreting  neutron star  binaries, matter  is channelled  along the
magnetic field lines  from the accretion disc to the  poles forming an
accretion  column~\cite{ghosh78,basko76}.  The  matter accumulated  at
the base of such accretion columns can significantly distort the local
magnetic  field due~\cite{melat01,payne04,mukhe12}.  Understanding the
evolution of the accretion mounds and its effect on the neutron star's
magnetic field will help address  several open questions, which can be
broadly classified into two categories:

\ben
   \i {\em Short term evolution of the accretion column:}
   The accreted matter  will be confined by the magnetic  field at the
   polar cap of the neutron star. However, beyond a threshold accreted
   mass, pressure driven plasma instabilities will cause the matter to
   break out of  the magnetic confinement and spread  over the neutron
   star~\cite{mukhe13a,mukhe13b,ferri13}.    Such  dynamic   processes
   occurring over short time scales  (local Aflven times and accretion
   time scales) will leave a significant imprint in the local magnetic
   field  topology.  The  dynamics  of the  spreading  of matter  from
   accretion  columns  remains  an   unresolved  question,  both  from
   observational and theoretical perspectives.

   The local field  topology can be directly  probed from observations
   of  cyclotron  resonance  scattering  features  (CRSF)  from  X-ray
   observations~\cite{hardi06}. Distortions  in the magnetic  field is
   also  expected  to  cause  to  complex  line  shapes  in  the  CRSF
   profiles~\cite{mukhe12,mukhe13a}.  Due  to limited  capabilities of
   existing  instruments, there  have been  only a  few detections  of
   asymmetric     line    profiles~\cite{potts05,furst15}.      Future
   observations with Astrosat and NuStar  is expected to constrain the
   physics of accretion columns further.  Theoretical understanding of
   the interplay of  radiation pressure and in-falling  matter and its
   effect  on continuum  radiation  is also  currently lacking.   Such
   studies  will help  better  constrain models  of accretion  columns
   which is currently used to model X-ray observations.
   \i {\em Long term evolution of the magnetic field:}
   The low magnetic field strengths  of millisecond pulsars have long
   been       attributed        to       recycling        due       to
   accretion~\cite{bisno74,roman90}.  However, the  exact mechanism of
   field decay  is still unclear.  Although  diamagnetic screening due
   to            field            burial           has            been
   proposed~\cite{melat01,payne04,choud02,konar04b}   as  a   possible
   mechanism, there is no clear consensus on whether large scale field
   burial is  possible without  re-emergence~\cite{cummi01} or  if MHD
   instabilities~\cite{mukhe13a,mukhe13b}  may inhibit  deformation of
   field  lines  required  to   reduce  the  apparent  dipole  moment.
   Accretion      enhanced      ohmic      decay      of      magnetic
   fields~\cite{konar97c,konar97a,konar99b}   is   another   plausible
   mechanism that  may reduce the  magnetic fields to  values observed
   for millisecond pulsars.  However the  current works do not address
   in  detail  the  dynamics  of  spreading  resulting  from  magnetic
   channelling  of matter  onto the  neutron star,  which needs  to be
   addressed. Such studies will also  help understand the evolution of
   magnetic  field  strength of  accreting  neutron  stars from  $\sim
   10^{12}$ G to $\sim 10^{8}$ G in millisecond pulsars.

   Recently two accretion powered pulsars have shown evidences of long
   term evolution of the magnetic  field through evolution of the CRSF
   centroid energy. Her X-1 has shown  long term evolution of the CRSF
   independent to other observables both in  the form of a sudden jump
   followed  by a  gradual decay~\cite{staub14,kloch15}.   4U 1538-522
   also shows signatures of an evolving CRSF energy~\cite{hemph16}.
\een

\section{The Interior}

\subsection{State of the matter}
\label{eosns}

Shortly after the  discovery of pulsars, the study of  dense matter in
the  core of  neutron stars  had gained  momentum~\cite{glend96}.  The
rapid accumulation of  data on compact stars in recent  years may shed
light on  the gross  properties of  cold dense  matter far  off normal
nuclear matter density.  Neutron star  matter encompasses a wide range
of densities, from  the density of iron nucleus at  the surface of the
star to several times normal nuclear matter density in the core. Since
the chemical potentials of nucleons  and leptons increase rapidly with
density in  the interior of  neutron stars, several novel  phases with
large  strangeness fraction  such as,  hyperon matter,  Bose-Einstein
condensates of strange mesons and quark matter may appear there. It is
to be noted that strange matter  typically makes the equation of state
(EoS) softer  resulting in  a smaller maximum  mass for  neutron stars
than that for the nuclear EoS~\cite{glend96}.

Observed  masses and  radii  of  neutron stars  are  direct probes  of
compositions  and EoS  of the  interior.  The  theoretical mass-radius
relationship of compact stars could be directly compared with measured
masses  and  radii  from   various  observations.   Consequently,  the
composition  and  EoS  of  dense  matter in  neutron  stars  might  be
constrained.  Neutron  star masses  have been  estimated to  very high
degree  of accuracy.  This  has been  possible because  post-Keplerian
parameters  such as  time  derivative of  orbital  period, advance  of
periastron, Shapiro delay,  Einstein time delay etc.  were measured in
several  binary pulsars.  Currently  the  accurately measured  highest
neutron star  mass is 2.01$\pm 0.04$  M$_{\odot}$~\cite{anton13}. This
puts a strong constraint on the  EoS of neutron star matter. Those EoS
which  can  not  satisfy  the  2  M$_{\odot}$  constraint,  are  ruled
out~\cite{banik14}.

Unlike  masses,  radii  of  neutron stars  have  not  been  accurately
measured  yet.   After the  discovery  of  highly relativistic  binary
systems such as  the double pulsar system, PSR  J00737-3039, for which
masses of both the pulsars are  known accurately, it was argued that a
precise measurement  of moment  of inertia ($I$)  of one  pulsar might
overcome  the  uncertainties  in  the determination  of  radius  ($R$)
because dimensionally $I\propto M R^2$~\cite{latti05}. In relativistic
binary  systems, higher  order post-Newtonian (PN)  effects could  be
measured.  Furthermore, the relativistic  spin-orbit (SO) coupling may
manifest  in  an   extra  advancement  of  periastron   above  the  PN
contributions   such  that   the  total   advance  of   periastron  is
$\dot{\omega}    =   \dot{\omega}_{1PN}    +   \dot{\omega}_{2PN}    +
\dot{\omega}_{SO}$  \cite{damou88}. The  SO contribution  has a  small
effect  and  could be  measured  when  it  is  comparable to  the  2PN
contribution.   The  measurement  of  the   SO  effect  leads  to  the
determination of  moment of inertia of  a pulsar in the  double pulsar
system~\cite{ska02,ska04}. With  the present  day timing  accuracy for
the pulsar A of PSR J0737-3039, the determination of moment of inertia
at 10 percent level would take about 20 years.

This situation  would change with  the advent of the  SKA. Substantial
advancement  in the  timing precision  is  expected to  come from  the
SKA. The  high precision timing  technique in the SKA  would determine
the moment of inertia of a pulsar earlier than that in the present day
scenario. The accurate determination of  masses and moments of inertia
of  pulsars in  relativistic  binary  systems with  the  SKA leads  to
simultaneous  knowledge about  masses, radii  and spin  frequencies of
pulsars which would  be used to confront  different theoretical models
and constrain  the EoS  and compositions in  neutron star  interior or
even  yield the  EoS  in  a model  independent  way  by inverting  the
Tolman-Oppenheimer-Volkoff   equation~\cite{lindb92}.   The   EoS  and
compositions of dense matter  extracted from neutron star observations
are  also important  for the  construction  of EoS  tables for  CC-SNe
simulations, neutron star mergers  and understanding the appearance of
strange   matter   in    the   early   post   bounce    phase   of   a
CC-SNe~\cite{banik14}.

The spin-off from the measurement of  moment of inertia in the SKA era
will be manifold. It was already  predicted that the plot of moment of
inertia  versus  rotational  velocity  ($\Omega$)  might  reveal  some
interesting features of  pulsars. It was shown that  after the initial
spin down of a pulsar along a supra-massive sequence, there was a spin
up   followed  by   another  spin   down  in   the  I   vs.   $\Omega$
plane~\cite{weber99,zduni04,banik04}.   This  is  known  as  the  back
bending (S-shaped curve in the  plot) phenomenon.  This phenomenon was
attributed to  the strong  first order  phase transition  from nuclear
matter to some  exotic (hyperon, kaon condensed or  quark) matter. The
SKA  might provide  an  opportunity to  investigate  the back  bending
phenomenon and the existence of exotic matter in pulsars.

Another interesting  possibility is the presence  of super-fluidity in
neutron star matter. Generally it is inferred that pulsar glitches are
the  manifestation  of super-fluid  neutron  matter  in neutron  stars
\cite{ander12}.  Recently,  it has been  argued whether the  moment of
inertia of the  super-fluid reservoir in the inner  crust is sufficient
to explain  the latest  observational data of  pulsar glitches  or not
\cite{ander12,chame13a}. When the entrainment effect which couples the
neutron super-fluid  with the  crust, is taken  into account,  a larger
angular   momentum  reservoir   is   needed   for  observed   glitches
\cite{ander12}.   Consequently,  the  required  super-fluid  moment  of
inertia exceeds that of the super-fluid crust. This indicates that some
part of  the super-fluid core  would contribute to pulsar  glitches. It
would be  worth investigating the  super-fluidity in neutron  stars in
general  and the  super-fluid  moment of  inertia  fraction for  pulsar
glitches in particular using the  precision pulsar timing technique of
the SKA.

The  Indian  Neutron  Star   Community  has  tremendous  expertise  in
theoretical  modelling  of the  EoS  of  dense  matter and  mass-radius
relationships  of  (non)rotating  neutron  stars and  will  contribute
immensely in the science program of the SKA. 

\subsection{EOS constraints from thermonuclear bursts}

Thermonuclear  X-ray bursts  are observed  from neutron  star low-mass
X-ray  binaries.  These  bursts originate  from intermittent  unstable
thermonuclear burning  of accumulated  accreted matter on  the neutron
star   surface~\cite{stroh06}.   Thermonuclear   bursts  provide   the
following methods  to measure  neutron star  parameters, and  hence to
constrain  the  theoretically proposed  equation  of  state models  of
neutron star cores (see \citeN{bhatts10} and references therein).  (1)
Continuum  spectrum method:  fitting of  the continuum  burst spectrum
with an  appropriate model can be  useful to measure the  neutron star
radius.  (2) Spectral line method:  atomic spectral line observed from
the surface of  a neutron star provides a clean  method to measure the
neutron star radius  to mass ratio~\cite{bhatts06}. However,  so far a
reliable detection of such a line has not been done.  (3) Photospheric
radius  expansion  burst  method:  a  strong  burst,  which  shows  an
expansion of the photosphere, can be used to constrain the mass-radius
space of  neutron stars~\cite{ozel06}.  (4) Burst  oscillation method:
intensity  variation during  thermonuclear X-ray  bursts, i.e.,  burst
oscillation, provides the neutron star spin frequency with an accuracy
usually  much   better  than   1\%~\cite{chakr03}.   The   fitting  of
phase-folded  burst  oscillation  light  curves  with  an  appropriate
relativistic  model can  be useful  to measure  neutron star  mass and
radius~\cite{bhatt05,lo13}.    (5)  Millihertz   (mHz)  quasi-periodic
oscillation (QPO)  method: mHz  QPO, which originates  from marginally
stable thermonuclear burning on neutron  stars, can be used to measure
the stellar  surface gravity, and  hence to constrain  the mass-radius
space   of   neutron   stars~\cite{heger07}.  Given   the   systematic
uncertainties in  measurements, a joint  application of some  of these
methods can be very useful to constrain neutron star parameters. Burst
properties can be  studied with current and  future X-ray instruments,
including those of the upcoming Astrosat.  In its time, only the LAXPC
instrument  of  Astrosat will  have  the  capability to  detect  burst
oscillations. The above mentioned methods will be complementary to the
capability of SKA to measure neutron star parameters~\cite{ska02}.

\section{The Magnetosphere}

\subsection{Radio Pulsars}

A majority of the known neutron stars are radio pulsars, and have been
detected via  their radio emission.   Soon after the discovery  of the
first pulsar \cite{hewis68}, it became clear that the pulsed nature of
the received  signal is likely due  to a misalignment of  the rotation
and the  magnetic axes of  the pulsar~\cite{radha69}. Yet,  a complete
understanding of  the underlying  physical mechanisms  responsible for
the radio emission is far from complete.

The key to  the pulsar puzzle lies in a  critical understanding of the
emission  region  geometry  through  a comparison  of  the  high  time
resolution pulse data and  high-sensitivity precision polarimetry with
quantitative theoretical predictions. One  of the limitations has been
the lack of  precision data. The radio emission region  of a pulsar is
small and it  is at an altitude of a  few hundred kilometres depending
on  field geometry.   It is  thought  that the  radio-loud regions  in
pulsar magnetosphere are ruled by plasma electrodynamics in a rotating
system. The  magnetic field is  strong enough to constrain  the plasma
flow  to one  dimension, and  quantise the  gyro-motion.  The  induced
electric field  is strong  enough to accelerate  charges to  very high
Lorentz  factors.  A  relativistic  plasma within  this  system  emits
coherent  radiation  as  a  by-product of  pair  creation  and  plasma
dynamics. Although this hypothesis has gained wide acceptance, it must
be tested  by measurements using  the widest possible  bandwidths, the
highest possible time resolution and  the best possible sensitivity of
the proposed SKA.   An extreme antithetical model is one  in which the
emission  is infinitely  beamed  radiating tangentially  to the  local
magnetic field lines~\cite{ganga04}.

The  sensitivities and  the  ranges of  operating  frequencies of  the
present  instruments  are such  that  these  are sufficient  only  for
certain specific  magnetospheric studies. But there  exist other areas
of investigation which  are possible only with  sufficient increase in
sensitivity  that will  be facilitated  by  SKA. A  subset of  studies
related to the pulsar emission mechanism, broadly divided on the basis
of the  timescales involved and  the extents of the  emission regions,
are discussed below.
\ben
   \i {\em Phenomena at single  pulse timescales:} There are primarily
   3 kinds of intriguing phenomena  observed in single pulse sequences
   of  a significant  number of  pulsars ---  pulse-nulling, sub-pulse
   drifting and  mode-changing. Nulling  is an  interesting phenomenon
   wherein the pulse suddenly disappears, implying possibly a complete
   cessation of emission  or emitted flux density  below the detection
   sensitivity  of  current  generation  telescopes.   Recent  studies
   exploiting   simultaneous    multi-frequency   observations   using
   different  telescopes  suggest   that  the  magnetospheric  changes
   responsible   for    observed   nulling    occur   at    a   global
   scale~\cite{gajja14}.  Underlying  periodicities and  clustering of
   nulls and bursts  in few strong pulsars indicate the  presence of a
   stochastic  Poisson   point  process~\cite{gjk12,gjw14}.  Detecting
   nulls with  duration of one  or only a  few pulse periods  has been
   possible only for bright pulsars.  The high sensitivity of SKA will
   help in  (1) detection  of any faint  emission during  the apparent
   nulls,   and  (2)   studying  the   nulling  properties   of  faint
   pulsars. The different bands available in SKA-low and SKA-mid would
   be very useful to  extend simultaneous multi-frequency observations
   of a much larger sample of pulsars.  The increased sensitivity will
   also help in  studying the emission properties of  pulsars when the
   transition from  one profile mode  to the other takes  place.  Both
   these aspects will provide crucial  inputs to modelling of physical
   theories explaining these phenomena.

   Sub-pulse  drifting  or  sub-pulse modulation  involves  intriguing
   modulations  in single  pulse  components,  which indicate  towards
   physical processes  occurring at a  range of timescales ---  from a
   few milliseconds to  a few hundreds of  seconds. A phenomenological
   model   (carousel  model)   to  explain   sub-pulse  drifting   was
   suggested~\cite{ruder75} at a very early stage.  The carousel model
   was    modified     to    address    some    of     the    observed
   inconsistencies~\cite{gil00}.   Significant observational  advances
   have also been made in this direction.  In some bright pulsars, the
   sub-pulse drifting has  been shown to be in phase  from 300 to 1400
   MHz~\cite{jos13a}. Several  pulsars have been studied  in detail,
   with some  studies providing strong  support to the  above carousel
   model~\cite{vivek99,deshp99,deshp01,asgek05}   while   the   others
   indicating        towards       inconsistencies        in       the
   model~\cite{edwar03,maan14a}.   Some other  models  also have  been
   proposed~\cite{cleme04,cleme08,jones13,jones14}   to  explain   the
   sub-pulse modulation in pulsars.  Further observational progress on
   this  front,  like systematic  tests  of  various proposed  models,
   requires  high  sensitivity   and  good  quality  full-polarimetric
   measurements  of  pulsars  at  wide  range  of  frequencies.   Such
   measurements are possible  with the existing telescopes  only for a
   sample of bright pulsars. SKA will make it possible to extend these
   studies to even the fainter pulsars, and hence, aid in developing a
   robust physical model applicable to majority of pulsars.
   \i   {\em  Phenomena   at  micro-   and  nano-second   timescales:}
   Micro-structures    (less-ordered    intensity   variations    with
   time-scales 1 to 500~$\mu$s) are seen in nearly all bright pulsars,
   but  no consensus  has  been  reached as  to  their  origins It  is
   suspected  that  these  are  tied to  plasma  dynamics  within  the
   emission regions.  Detection and study of micro-structures obviously
   needs detection of  single pulses with sufficiently  high signal to
   noise ratios (S/N).

   Giant pulse  (GP) emissions --  a phenomenon currently known  to be
   exhibited by only  about a dozen radio pulsars (out  of nearly 2500
   known)  --  are  short-duration  (sometimes   as  short  as  a  few
   nanoseconds)  burst-like  sporadic  increases of  individual  pulse
   intensities~\cite{hanki03}.   The peak  flux densities  of GPs  can
   exceed those of regular individual pulses by factors of hundreds or
   even thousands. Although several  mechanisms have been proposed for
   the observed  GP emission,  there is no  satisfactory answer  for a
   question  as   simple  as  following:  what   are  the  identifying
   characteristics of the pulsars that emit giant pulses?

    Large  bandwidths are  required  to  separate intrinsic  frequency
    structure   from  that   imposed   by   propagation  through   the
    inhomogeneous interstellar medium,  i.e., interstellar scattering.
    New  technology   in  the   form  of   low-noise  decade-bandwidth
    dual-polarisation   antenna   feeds   and   associated   low-noise
    amplifiers are  required as  well as  data recording  systems fast
    enough  to sample  these  bandwidths.   The increased  sensitivity
    facilitated  by   SKA  will  help   in  detection  and   study  of
    micro-structures as well as giant pulses from a statistically large
    number of pulsars. The purity  of the polarisation measurements of
    such detections will  also provide crucial help  in localising the
    physical  emission  regions,  and   hence,  in  understanding  the
    emission mechanism of these features.
   \i   {\em   Ultra-wide   bandwidth,  ultra-high   time   resolution
     observations} are  critical, because models diverge  in what they
    predict for short time  scales and fundamental emitter bandwidths.
     The emission changes within one  rotation period, so we must have
     the  {\em   highest  possible  sensitivity}   to  see  individual
     ``pulses''.  We need  radio  observations which  can resolve  the
     dynamic time-scales of the plasma (on the order of $\sim 10^{-7}$
     to $10^{-5}$  s), the intrinsic plasma-turbulent  time scales (as
     short as  $\sim 10^{-9}$ s), and reveal  the intrinsic bandwidths
     of the emission.
   \i {\em Motion of emission point:} With the advent of long baseline
   interferometry it is possible  to measure the astrometric motion of
   pulsar emission point  with respect to rotation phase~\cite{pen14}.
   The relativistic  effects such as aberration  and retardation (A/R)
   effects   indeed   change   the   locations   of   emission   point
   coordinates~\cite{ganga01,gupta03,ganga05},   and  they   are  much
   effective in  millisecond pulsars compared to the  normal ones.  To
   resolve  the emission  region of  a pulsar  would require  nano- or
   picoarcsecond  imaging, which  is challenging  to achieve  with the
   existing telescopes.
   \i  {\em   Polarimetry:}  Resolution   of  single  pulses   to  the
   micro-structure level with full  Stokes polarisation is required to
   advance our  understanding of  the pulsar radio  emission mechanism
   and  the propagation  effects  in the  pulsar magnetosphere.   Wave
   polarisation  provides almost  all information  about the  emission
   geometry  and   reflects  the   physics  of  the   emission  and/or
   propagation directly.  High sensitivity is absolutely a key because
   useful  polarimetry requires  that  the received  signal level  $S$
   substantially exceed the  noise level $N,$ e.g.,  $S/N\gg $1.  What
   determines the linear  and circular polarisation of  a signal? What
   causes   the   rapid   orthogonal  mode   transitions   in   linear
   polarisation, and the rapid  sign changes of circular polarisation?
   Are  these a  signature of  the emission  process, or  a result  of
   propagation  in  the  pulsar magnetosphere?   Although  calibration
   techniques for  accurate polarimetry are now  well known, attention
   must be  paid to the  polarisation characteristics of  new wide-band
   feed  systems   to  assure   that  they   can  be   accurately  and
   unambiguously calibrated.  To take advantage of these wide bands we
   will need  fast digital  ``backend'' data acquisition  systems with
   high  dynamic   range  to   allow  interference   excision  without
   corrupting pulsar signals in interference-free bands.
   \i {\em Pulsar Emission Physics:} The SKA is expected to greatly help
   us to address some of the key problems in pulsar physics.
   \bei
      \i {\em How can we construct the 3D structure of pulsar emission
        beam?} Currently  the 3D structure  of pulsar emission  is not
      clear whether  it is conal  or patchy.  There are  arguments for
      concentric   rings   near   and  around   the   magnetic   field
      axis~\cite{ranki83,ranki93,ganga01,mitra02} or  random locations
      in   the    beam   rather   than   in    some   coherent   conal
      structure~\cite{lyne88b}.   \citeN{ganga01} have  further showed
      that  at  any  given  radio frequency  the  emissions  close  to
      magnetic axis  comes from lower  altitude compared to  the conal
      emissions. The  individual pulses that build  the stable profile
      show enormous  diversity in  their overall  characteristics. One
      needs to understand  how the single pulses averaged  to a stable
      mean profile.   Deduction of  3D structure  of the  pulsar radio
      beam and how it forms  from individual pulses, comprises a major
      step in revealing the pulsar  radio emission mechanism.  One has
      to study  in a  large survey  the beam shapes  of young  and old
      pulsars to resolve this issue.
      \i {\em Modelling of polarisation of pulsar profiles -} Both the
      emission process  and the viewing geometry  set the polarisation
      state  of the  pulsar radio  emission, which  is expected  to be
      further   modified  by   the  propagation   effects  in   pulsar
      magnetosphere.   \citeN{ganga10}  has developed  a  polarisation
      model of pulsars based on  curvature emission, and shown how the
      S-type  polarisation angle  swing correlates  with the  sense of
      circular  polarisation.  Further,  Kumar  \& Gangadhara  (2012a,
      2012b, 2013)~\nocite{kumar12a,kumar12b,kumar13} have generalised
      the  polarisation model  to  include aberration-retardation  and
      polar cap currents.  Propagation effects have been observed, the
      most prominent being orthogonal modes of polarisation, which are
      generated  in  the  magnetospheric  plasma.   Understanding  the
      nature   and  origin   of  orthogonal   polarisation  modes   in
      magnetospheric  plasma   is  crucial  in   understanding  pulsar
      emission physics.  The observing capabilities of SKA is expected
      to resolve  the issue of  the origin of  orthogonal polarisation
      modes: intrinsic  to the  emission process  or generated  by the
      propagation effects.
   \eei
   \i  {\em Continuous/unpulsed/off-pulse  Emission :}  Presence of  a
   continuous  emission component,  i.e.,  emission  in the  off-pulse
   regime or  far from the  main pulse in  the profile, has  been long
   looked for. It is only recently  that such emissions have been made
   from  the  pulsars   B0525$+$21  and  B2045$-$16~\cite{basu11}.   A
   magnetospheric origin  of such off-pulse emission  raises questions
   about the  location of the  emission region. These  detections have
   been  possible due  to the  availability  of a  fast sampling  time
   ($\sim$ 125~ms)  in the interferometric  mode of the  GMRT.  GMRT's
   fast sampling  time is adequate  to resolve the  off-pulse emission
   only  for  a handful  of  pulsars,  that  too very  coarsely.   The
   strength of the  off-pulse emission has also been found  to be only
   about  $1\%$ of  that  of the  main pulse,  demanding  a very  high
   sensitivity  instrument.   The  SKA,  with  its  high  sensitivity,
   possible gating  in 100\,bins  across the  pulsar period  and large
   frequency coverage will  be an ideal and much  awaited telescope to
   carry out the off-pulse emission searches and studies.
\een

Apart from  the known emission  components detailed above,  very faint
radio  emission  from  a  some   of  the  gamma-ray  pulsars,  earlier
considered to be {\em radio-quiet},  have been detected.  Two of these
gamma-ray pulsars, J0106+4855 and J1907+0602,  have been found to emit
in      radio      with      L-band     flux      densities      below
10\,$\mu$Jy~\cite{plets12,abdo10}.   Furthermore,   another  gamma-ray
pulsar,  J1732$-$3131,   has  been   found  to   emit  at   low  radio
frequencies~\cite{maan12,maan14b,maan16}  with an  upper limit  on its
flux density at L-band being 50\,$\mu$Jy.  Unusual radio emission from
these  gamma-ray  pulsars  might  also  be detected  in  the  form  of
``giant-pulses'' or bursty  emission~\cite{maan15}.  Detection of such
faint  or  bursty radio  emission  from  these pulsars  might  suggest
presence  of radio  emission from  all the  gamma-ray pulsars  that is
below the  detection sensitivity of current  telescopes.  The upcoming
telescopes SKA and FAST, with their unprecedented sensitivities, might
uncover such faint emission from these pulsars, and hence, a new class
of  faint  radio  pulsars.  Detection   of  the  faint  emission  with
polarimetric   information  will   also   play  a   crucial  role   in
understanding the location of  the gamma-ray emission regions relative
to the radio emission regions.

\subsection{Fast Radio Bursts}

Fast      radio       bursts      (FRB)      are       a      recently
discovered~\cite{lorim07b,thorn13,spitl14b,ravi15}   class   of   radio
transients which are of very short duration ($\sim$ millisecond), show
characteristics  of  propagation  through  cold, diffuse  plasma,  and
likely  originate  at  cosmological  distances.  There  are  different
hypotheses for creation  mechanism of FRBs, including super-conducting
strings   \cite{vacha08,yu14},   merger   of   binary   white   dwarfs
\cite{kashi13}   or   neutron   stars  \cite{totan13},   collapse   of
supra-massive  neutron  stars  \cite{falck14},  exploding  black  holes
\cite{barra14},  dark   matter  induced  collapse   of  neutron  stars
\cite{fulle15},  and many  others.  None  of the  above  hypotheses is
established beyond doubt, and the understanding of the physical origin
of FRBs remains as an open challenge. Moreover, discovery of more FRBs
might  lead  to  better  understanding  of  the  intergalactic  medium
\cite{zheng14}.

A theoretical understanding of FRBs,  supplemented by a search for new
FRBs in existing pulsars  surveys, is of significant recent interest.
Already, the Parkes FRB triggers  are being investigated with the GMRT
for their  afterglow emissions. Moreover, there is  an ongoing project
to develop  a transient detection system at  GMRT \cite{bhat13}, which
might  be very successful  to detect  FRBs.  More  details on  FRBs and
description of  activities and interests among  Indian researchers can
be found in  the write-up presented by the  `Transient Science Working
Group'.

\section{Gravitational Physics : Pulsar probes}

Two  key Science  goals of  the SKA  involve exploring  the nature  of
relativistic gravity  and to directly detect  nano-Hertz gravitational
waves, predicted in general relativity.  At present, Pulsars in binary
systems are extremely successful in  testing general relativity in the
strong    field    regime    \cite{stair03,stair04,krame06,stair10}. 
These pulsar binaries  usually include neutron star-white
dwarf  (NS-WD)   and  neutron   star-neutron  star   (NS-NS)  systems.
Unfortunately, neutron star-black hole (NS-BH)  binaries are yet to be
discovered, although  different studies  on such possible  binaries in
the Galactic disk \cite{pfahl05,kiel09}, globular clusters \cite{sigur03},
and  near the  Galactic centre  \cite{fauch11} have  been presented.   If
NS-BH binaries have small orbital periods (around a day), as predicted
\cite{pfahl05,kiel09}, these  might lead  to superior  tests of  general
relativity,  provided technical  difficulties can  be overcome.  These
systems might also help to determine  the spin parameter of the BH and
test the  validity of the {\it  Cosmic Censorship Conjecture }  and to
test the {\it  BH no-hair theorem} \cite{ska04}.  As  NS-BH binaries are
also  important sources  for  gravitational waves  for Advanced  LIGO,
understanding  of the  properties of  these systems  from pulsar  data
analysis will  help the gravitational  wave community to  build better
waveform templates.

The  orbital  dynamics of  pulsars  in  binary systems  are  generally
described  in  terms  of   five  Keplerian  and  eight  post-Keplerian
parameters \cite{damou86,kopei94,lorim04}. The leading order expressions
under  general  relativity  have  been  used  for  the  post-Keplerian
parameters. Measurements  of these post-Keplerian  parameters (through
pulsar timing analysis) lead to the determination of the masses of the
pulsar and  the companion.  For  a NS-BH  binary, the values  of these
post-Keplerian  parameters will  be  larger, e.g.,  the Shapiro  range
parameter for a NS-BH binary is more than seven times larger than that
for  a NS-NS  binary \cite{bagch14}.  Such high  values of  these leading
order  post-Keplerian parameters  for NS-BH  systems imply  that these
terms will  be measurable  even with a  shorter data  span.  Moreover,
even the higher  order terms might be significant, and  if that is the
case, one  would need  to incorporate these  higher order  terms while
performing  timing  analysis  to  avoid  obtaining  inaccurate  system
parameters \cite{bagch13,bagch14}.

The central  region of the  galaxy has  a dense population  of visible
stars.  It is  quite likely that compact objects, such  as black holes
(BHs) \cite{morri93,freit06b},  neutron stars and Intermediate  Mass Black
Holes (IMBH)  \cite{porte06} may also  be present there.   The discovery
and  timing  of  millisecond  pulsars  in the  centre  of  our  galaxy
(hereafter GC)  may allow us  to detect long  wavelength gravitational
waves  (GWs)  emitted  from  the  SGR A*  region  due  to  large  mass
black-holes orbiting  the central  super-massive black-hole  (SMBH). This
will allow us to ``gravitationally  probe" these crowded regions which
are    usually    obscured     in    the    electromagnetic    channels
\cite{kocsi12}.  The  GW signal generated by  a population
of objects  (the ``foreground")  is smooth if  the average  number per
$\Delta f$  frequency bin satisfies  $\langle \Delta N\rangle  \gg 1$.
The GW spectrum becomes spiky  (with $\langle \Delta N\rangle \leq 1$)
above a critical frequency $f_{\rm res}$ that depends on the number of
objects  within 1pc  of the  GC and  on the  timing observation  span.
Sources within  $r_{\rm res}$  generate distinct spectral  peaks above
frequency $f_{\rm res}$.  These sources  are {\it resolvable}.  The GW
spectrum transitions from continuous to discrete at higher frequencies
inside  the  Pulsar  Timing  Array frequency  band.   If  pulsars  are
observed repeatedly  in time for an  observation span $T =  10$ yr and
with an interval $\Delta  t = 1$ week, this can probe  the range of GW
frequencies: $3  \times 10^{-9} \; \rm  Hz \; (3  nHz) < f <  3 \times
10^{-6} \; \rm Hz \; (3000 nHz)$.  The cosmological GW background from
the  whole  population  of  massive black  hole  binaries  (MBHBs)  is
actually an  astrophysical ``noise" for  the purpose of  measuring the
GWs  of  objects orbiting  SgrA*.   The  characteristic GW  amplitudes
(either of a  stochastic background or of a resolvable  source) can be
translated into  into a ``characteristic timing  residual" $\delta t_c
(f)$ corresponding to a delay in the time of arrivals of pulses due to
GWs, after  averaging over  the sky  position and  polarisation.  The
results of simulations  are summarised in Fig.\ref{fig:spectral_index}.
BHs  in orbit around
SMBH SgrA*  generates a continuous GW  spectrum with $f <  40$ nHz.  A
100 ns  - 10  $\mu$s timing  accuracy with SKA  will be  sufficient to
detect  IMBHs  ($1000  \;  M_{\odot}$),  if they  exist,  in  a  3  yr
observation if stable PSRs $0.1- 1$ pc away from SgrA* are timed.

\bef
\begin{center}
\includegraphics[width=0.75\textwidth]{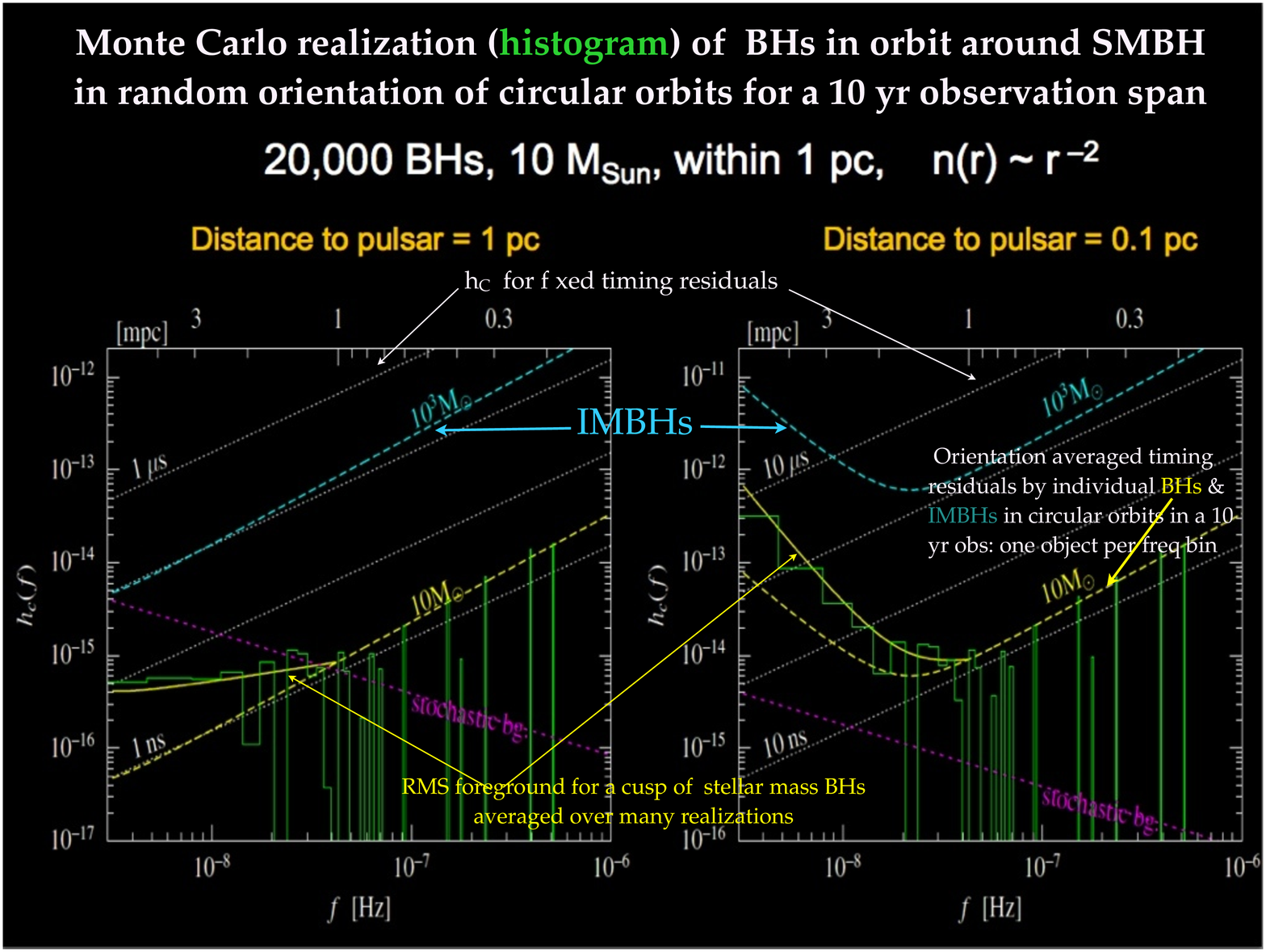}
\caption{Detectable  phase space  of  characteristic strain  amplitude
  $h_c$ and GW frequency $f$ with pulsar timing with a probe pulsar at
  1\,pc (left) and  0.1\,pc (right) from the  Galactic Centre.  Dotted
  white  lines show  the orientation-averaged  $h_c$ for  fixed timing
  residuals measured from  the probe pulsars.  Yellow  and cyan dashed
  lines  show  respectively  the  binary  orientation-averaged  timing
  residuals caused  by individual  stellar BHs  and IMBHs  on circular
  orbits  around the  SMBH in  a  10 yr  timing of  the probe  pulsar,
  assuming  1   object  (source)  per  frequency   bin.   Green  lines
  (histogram) show  the timing residuals  for a random  realisation of
  $10 \; \rm M_{\odot}$ stellar BHs  in the cluster (20,000 BHs within
  1 pc with  number density $\propto r^{-2}$).  At high  $f$, only few
  bins are  occupied, generating a  spiky signal.  At lower  $f$, many
  sources  overlap to  create a  continuous spectrum.   Magenta dashed
  lines show  the cosmological stochastic  GW background which  can be
  much smaller, especially  as $f$ increases.  For  other (steeper) BH
  population density profiles or for  BH orbits with isotropic thermal
  eccentricity distributions, see Kocsis et al (2012).  }
\label{fig:spectral_index}
\end{center}
\eef

A  large  population  of  pulsars  could  be  present  inside  the  GC
(\cite{pfahl04}).  The recent  discovery of GC magnetar  SGR J1745-29 in
the X-ray bands  with NuSTAR and subsequently in the  $1.2- 18.95$ GHz
radio bands (\cite{bower14}, \cite{spitl14a}) shows that the source angular
sizes  are consistent  with scatter  broadened size  of SgrA*  at each
radio  frequency. Additionally,  pulse broadening  timescale at  1 GHz
(\cite{spitl14a})  is   several  orders  of  magnitude   lower  than  the
scattering  predicted  by  NE2001  model  \cite{corde02}.   \cite{chenn14}
estimate an  upper limit  of $\sim  950$ potentially  observable radio
loud  pulsars in  GC.  However,  \cite{dexte14} point  out that  despite
several deep radio surveys no ordinary pulsars have been detected very
close  to the  GC and  suggest an  intrinsic deficit  in the  ordinary
(i.e. slow) pulsar population.   \cite{macqu15} distinguish two possible
scattering  scenarios  affecting  the search  for  millisecond  pulsar
search in  the GC and  suggest that in  the weak scattering  regime (if
applicable for a  large part of the GC) a  substantial fraction of the
pulsars ($> 1  \; \rm mJy\; \rm  kpc^2$ at 1.4 GHz)  would be detected
when observed  with SKA-Mid  in the  X-band ($\sim  8 \;\rm  GHz$) and
possibly a  smaller fraction  even with  EVLA and GBT  at a  less deep
level.  For  the  strong  scattering regime  however,  the  full  high
frequency ($\sim  25 \;\rm GHz$) capability  of the SKA on  the longer
term will be  necessary.  The detection and long  term accurate timing
of stable millisecond  pulsars near the GC will be  of interest to the
Indian partners  of the  SKA community  from multiple  perspectives of
probing the  contents of the galactic  centre, detecting gravitational
waves and tests of strong gravity.

A  NS-BH system  will be  a very  good tool  to test  the validity  of
non-conservative   theories  of   gravity,   which   produce  a   self
acceleration of the centre of  mass of a binary \cite{bagch14}. Moreover,
within the framework of general  relativity and many other theories of
gravity,  any perturbation  in the  space-time (like  rotating neutron
stars  or black  holes, neutron  stars with  other compact  objects as
binary   companions,  etc.)   produces  ripples,   i.e.  gravitational
waves.  For  the case  of  general  relativity,  the emission  of  the
mono-polar gravitational wave is forbidden  by the conservation of mass
and the emission of the dipolar gravitational wave is forbidden by the
conservation  of momentum.  The  quadrupolar emission  remains as  the
lowest  order  mode of  emission  of  gravitational waves  under  this
theory.  The  existence  of  such emission  has  been  established  by
measurement of  the decrease of the  orbital period of many  NS-NS and
NS-WD binaries.  But there are many  hypothetical alternative theories
of gravity (like the `scalar-tensor' theories) which allow emission of
mono-polar  and  dipolar gravitational  waves.  NS-BH  systems will  be
better  systems to  detect such  emissions as  the combined  effect of
mono-polar, dipolar and quadrupolar  gravitational wave emission is much
larger for such  a system than that of NS-WD  systems \cite{bagch14}.  It
will be interesting  to probe the implications  of employing adiabatic
precessional equation for the orbital  plane in the context of testing
the {\it BH no-hair theorem}.  It  turns out that going beyond such an
adiabatic approximation  is relevant while  constructing gravitational
wave  inspiral templates  \cite{gopak11,gupta14}.   Going  beyond the  above
adiabatic approximation can lead  to certain quasi-periodic variations
of angles  that specify the  orbit.  This is qualitatively  similar to
quasi-periodic   evolution   of   few  Keplerian   parameters   while
incorporating  the   effect  of  gravitational  wave   emission  in  a
non-adiabatic manner as detailed in \cite{damou04}.

However, a NS-BH  system will not be  as good as a  NS-WD system while
trying to probe  possible variation in the value  of the gravitational
constant  $G$ or  to  test strong  equivalence principle  \cite{bagch14}.
Another interesting possibility will  be to observe millisecond pulsar
binaries whose companions are visible in the optical and near-infrared
wavelengths with large  telescopes like the Thirty  Meter Telescope in
the  square  Kilometre array  era.   The  combined optical  and  radio
observations of  such double spectroscopic binaries  should eventually
allow  us  to  test  `scalar-tensor' relativity  theories  in  certain
interesting  regime   which  are  at  present   difficult  to  achieve
\cite{kharg12}.

On  the gravitational wave  aspect, we  plan to  pursue investigations
that  can provide  constructs,  relevant for  analysing pulsar  timing
array data, that model gravitational waves from massive spinning black
hole  binaries in  post-Newtonian eccentric  (and  hyperbolic) orbits.
These    investigations   are   expected    to   be    influenced   by
\cite{damou04,tessm07,gopak11,vitto14}  that   provided  accurate  and
effect prescription to construct post-Newtonian accurate gravitational
wave templates compact binaries in non-circular orbits.

Super-massive  Black Hole  binaries  (SMBHBs) are  putative sources  in
nano-Hertz  gravitational wave  astronomy.  Incoherent  superposition a
large number  of SMBHBs is  expected to produce  stochastic background
signal  for Pulsar  Timing  Arrays  (PTA). Though  it  is unlikely  to
resolve the sources individually, such a possibility has been explored
for z<2~\cite{sesan08,olmez10,ravi12}.  The authors further  study and
show that with 100 pulsars, the  SMBHBs can be located with 40 degree
square~\cite{sesan10}.  Pertaining to  the early  inspiral phase,  the
signal  from  such source  is  monochromatic  in nature.  The  maximum
likelihood  approach  has  been   further  developed  to  explore  the
feasibility to  localise the  SMBHB source with  a PTA~\cite{babak12}.
Recently, we have defined the figures  of merit of PTA to quantify its
efficiency and  probe the  angular resolution ability  as well  as the
polarisation      recovery      of      the      underlying      SMBHB
source~\cite{agarw16}.  This work  clearly demonstrates  the idea  of
using PTA  as a multi-detector  network to detect  gravitational waves
from SMBHBs.

The fast spinning accreting neutron  stars with an accretion mound are
potential   sources   for   the   ground  based   gravitational   wave
detectors~\cite{bilds98}.  These sources are attractive due their spin
frequency being several hundred hertz,  a match with the GW detectors.
At the  same time,  gravitational wave search  from these  sources are
very computation intensive because its detection will require coherent
analysis of  data over a long period  of time, a couple  of years. One
needs to know the spin and  orbital parameters of the system and their
evolution  very accurately  over  this entire  period. Otherwise,  the
parameters space for search is very large~\cite{watts09a} rendering it
impossible and  it will also  reduce the significance of  any detected
signal. The spin and orbital  evolution of a few accreting millisecond
pulsars are known~\cite{hartm08}, but unfortunately, these systems are
transient, and  therefore have lower average mass  accretion rate, aka
weak  gravitation  wave  signal.  Some  transient  sources,  like  EXO
0748-676, can however, be very  potential candidate as they spend long
periods in  X-ray bright state and  go into quiescence  in between. If
SKA finds  radio pulsations from  some of these sources  in quiescence
(same as the  the transitional LMXBS), which will  also help establish
the  orbital parameters, gravitation  wave searches  can be  done over
relatively smaller  parameter ranges during their  future X-ray bright
states.

\section{Multi-wavelength Studies}

Observations with  Astrosat will improve our  understanding of neutron
stars in  X-ray binaries in  several ways. The biggest  advantage that
Astrosat provides  is in  terms of large  effective area of  the LAXPC
instrument,  especially over  a  wide energy  band  extending upto  80
keV.  Among the  known and  yet to  be discovered  LMXBs,  Astrosat is
likely  to  discover more  accreting  millisecond  X-ray pulsars,  the
improved statistics  will give  better understanding of  the mechanism
and   limits  of   the   neutron  star   spin-up   via  accretion   in
LMXBs.  Discovery of more  pulsars in  the process  of spinning  up in
LMXBs, like the 11 Hz  accreting pulsar in the globular cluster Terzan
5 will be additional support for the process of spinning up of neutron
stars   through   their  LMXB   phase.    Observations   of  the   KHz
quasi-periodic  oscillations,  thermonuclear  burst oscillations,  and
thermonuclear burst  spectroscopy are also likely  to be significantly
improved   with  Astrosat.   All   of  these   are  very   useful  for
understanding the EOS of neutron stars.

Astrosat will  carry out  a lot  of study of  the high  magnetic field
accreting  neutron  stars,  more  commonly known  as  accreting  X-ray
pulsars, most of which are found in HMXB systems. Different aspects of
X-ray pulsar studies  with Astrosat that are of  wider interest are i)
magnetic  field  configuration  of  the neutron  stars,  ii)  possible
alignment of the spin and magnetic axis, iii) magnetic field evolution
in  the accretion  phase.  Understanding of  these  will improve  with
Astrosat measurements of energy  and intensity dependence of the pulse
profiles  of  these systems  and  pulse  phase dependence,  luminosity
dependence, and  time dependence of  the cyclotron line  parameters of
the X-ray pulsars. A type  of relatively newly known accreting systems
are the fast X-ray  transients with super-giant companion stars. Though
most  of these  systems are  expected to  harbour high  magnetic field
neutron stars, persistent X-ray  pulsations have been detected in only
three of  these systems and cyclotron  line has been  detected in only
one system. Astrosat observations are likely to bring more clarity to
the nature of the compact objects in these systems and perhaps provide
insight onto why  accretion from wind in these  systems give different
X-ray features in the form of fast transient outbursts.

Though the immediate emphasis of this  document has been the impact of
SKA  era  on neutron  star  research  and  the immediate  benefits  of
Astrosat,  we  need   to  remember  that  a  large   number  of  other
high-sensitivity instruments  (both in radio and  higher energies) are
also upcoming;  many of which  has active Indian  participation (like,
SKA,  TMT, LIGO  etc.).  It  is imperative  that maximal  advantage is
taken of the  science capabilities of these.   In tables [\ref{tab02}]
\& [\ref{tab03}] we  present a comprehensive list  of such instruments
and  note down  the particular  kind of  investigations that  could be
undertaken using them.

\section{Summary}

The  SKA  represents  the  next  big leap  in  sensitivity  for  radio
astronomy.  The neutron star science with  the SKA will mainly be done
through all-sky surveys,  which is expected to result  in tripling the
current pulsar population, allowing variety  of theories to be tested.
It is noted that the interests  of the neutron star community in India
correspond closely to these SKA science goals, in regard to the pulsar
astronomy.    We    are   now   beginning   to    define   theoretical
calculations/simulations as well as observational projects, keeping in
mind  that the  community can  immediately  make use  of the  recently
launched Indian X-ray instrument {\bf  Astrosat} and the upgraded GMRT
({\bf uGMRT})  which has been  given the {\bf SKA  pathfinder} status.
Execution  of these  projects, theoretical  as well  as observational,
would  prepare  the community  to  make  appropriate  use of  the  SKA
capabilities in future.

\section*{Acknowledgments}

The first  author, SK, is  supported by a grant (SR/WOS-A/PM-1038/2014)
from the Department of Science \& Technology, Government of India.

\bibliography{mnrasmnemonic,adsrefs}
\bibliographystyle{mnras}

\begin{deluxetable}{lllccc}
\tabletypesize{\footnotesize}
\tablecolumns{6}
\tablewidth{0pt}
\tablecaption{Pulsars with no associated SNRs}
\startdata
\hline
& Name          & Association                  & $\tau_s$  & {\bb}$_{\rm s}$ & {\edot}$_{\rm rot}$ \\ 
&               &                              &  Yr       &  G            & erg.s$^{-1}$ \\
\hline
  1 &  J1050-5953   &  XRS                         &  2.68e+03 & 5.02e+14 & 5.6e+33 \\ 
  2 &  J1023-5746   &  GRS(F)                      &  4.6e+03  & 6.62e+12 & 1.1e+37 \\ 
  3 &  J1838-0537   &  *                           &  4.89e+03 & 8.39e+12 & 6.0e+36 \\ 
  4 &  J0100-7211   &  EXGAL:SMC,XRS               &  6.76e+03 & 3.93e+14 & 1.4e+33 \\ 
  5 &  J1357-6429   &  XRS:PWN                     &  7.31e+03 & 7.83e+12 & 3.1e+36 \\ 
  6 &  B1610-50     &  *                           &  7.42e+03 & 1.08e+13 & 1.6e+36 \\ 
  7 &  J1617-5055   &  GRS                         &  8.13e+03 & 3.1e+12  & 1.6e+37 \\ 
  8 &  J1734-3333   &  *                           &  8.13e+03 & 5.22e+13 & 5.6e+34 \\ 
  9 &  J1708-4008   &  XRS                         &  8.9e+03  & 4.7e+14  & 5.8e+32 \\ 
 10 &  J1418-6058   &  GRS(F),GRS(H)               &  1.03e+04 & 4.38e+12 & 4.9e+36  \\
 11 &  J1301-6305   &  *                           &  1.1e+04  & 7.1e+12  & 1.7e+36  \\
 12 &  J1809-1943   &  XRS(AXP)                    &  1.13e+04 & 2.1e+14  & 1.8e+33  \\
 13 &  J1746-2850   &  *                           &  1.27e+04 & 3.85e+13 & 4.2e+34  \\
 14 &  J1420-6048   &  GRS(F),GRS(H)               &  1.3e+04  & 2.41e+12 & 1.0e+37  \\
 15 &  J1413-6141   &  *                           &  1.36e+04 & 9.88e+12 & 5.6e+35  \\
 16 &  J1826-1256   &  GRS(F),XRS(AXP),PWN         &  1.44e+04 & 3.7e+12  & 3.6e+36  \\
 17 &  J1702-4310   &  *                           &  1.7e+04  & 7.42e+12 & 6.3e+35  \\
 18 &  J2021+3651   &  GRS,GRS(F)                  &  1.72e+04 & 3.19e+12 & 3.4e+36  \\
 19 &  J2111+4606   &  GRS(F)                      &  1.75e+04 & 4.81e+12 & 1.4e+36  \\
 20 &  J2004+3429   &  *                           &  1.85e+04 & 7.14e+12 & 5.8e+35  \\
 21 &  B1046-58     &  GRS(F)                      &  2.03e+04 & 3.49e+12 & 2.0e+36  \\
 22 &  B1737-30     &  *                           &  2.06e+04 & 1.7e+13  & 8.2e+34  \\
 23 &  J1856+0245   &  GRS(H),XRS (AXP)            &  2.06e+04 & 2.27e+12 & 4.6e+36  \\
 24 &  J1935+2025   &  *                           &  2.09e+04 & 2.23e+12 & 4.7e+36  \\
 25 &  B1823-13     &  GRS,XRS:PWN                 &  2.14e+04 & 2.8e+12  & 2.8e+36  \\
 26 &  J1934+2352   &  *                           &  2.16e+04 & 4.89e+12 & 9.1e+35  \\
 27 &  J1958+2846   &  GRS(F)                      &  2.17e+04 & 7.94e+12 & 3.4e+35  \\
 28 &  J1838-0655   &  XRS (AXP), GRS(H)           &  2.27e+04 & 1.89e+12 & 5.5e+36  \\
 29 &  J1135-6055   &  *                           &  2.3e+04  & 3.05e+12 & 2.1e+36  \\
 30 &  J1909+0749   &  *                           &  2.47e+04 & 6.07e+12 & 4.5e+35  \\
 31 &  J1410-6132   &  *                           &  2.48e+04 & 1.28e+12 & 1.0e+37  \\
 32 &  J1747-2958   &  GRS (PWN)                   &  2.55e+04 & 2.49e+12 & 2.5e+36  \\
 33 &  B1727-33     &  *                           &  2.6e+04  & 3.48e+12 & 1.2e+36  \\
 34 &  J2238+5903   &  GRS(F)                      &  2.66e+04 & 4.02e+12 & 8.9e+35  \\
 35 &  J1821-1419   &  *                           &  2.93e+04 & 3.89e+13 & 7.8e+33  \\
 36 &  J1841-0524   &  *                           &  3.02e+04 & 1.03e+13 & 1.0e+35  \\
 37 &  J1524-5625   &  *                           &  3.18e+04 & 1.77e+12 & 3.2e+36  \\
 38 &  J1112-6103   &  *                           &  3.27e+04 & 1.45e+12 & 4.5e+36  \\
 39 &  J1718-3718   &  *                           &  3.32e+04 & 7.47e+13 & 1.7e+33  \\
 40 &  J1837-0604   &  *                           &  3.38e+04 & 2.11e+12 & 2.0e+36  \\
 41 &  J1833-0831   &  XRS (SGR)                   &  3.49e+04 & 1.63e+14 & 3.1e+32  \\
 42 &  J0729-1448   &  *                           &  3.52e+04 & 5.4e+12  & 2.8e+35  \\
 43 &  J1932+1916   &  *                           &  3.54e+04 & 4.46e+12 & 4.1e+35  \\
 44 &  J1551-5310   &  *                           &  3.68e+04 & 9.52e+12 & 8.3e+34  \\
 45 &  J1907+0918   &  *                           &  3.8e+04  & 4.67e+12 & 3.2e+35  \\
 46 &  J1015-5719   &  *                           &  3.86e+04 & 2.87e+12 & 8.3e+35  \\
 47 &  B1930+22     &  *                           &  3.98e+04 & 2.92e+12 & 7.5e+35  \\
 48 &  J1044-5737   &  GRS(F)                      &  4.03e+04 & 2.79e+12 & 8.0e+35  \\
 49 &  J1815-1738   &  *                           &  4.04e+04 & 3.98e+12 & 3.9e+35  \\
 50 &  J1637-4642   &  *                           &  4.12e+04 & 3.06e+12 & 6.4e+35  \\
 51 &  J1849-0001   &  XRS,GRS,GRS(H)              &  4.29e+04 & 7.49e+11 & 9.8e+36  \\
 52 &  J1745-2900   &  XRS (AXP)                   &  4.31e+04 & 7.3e+13  & 1.0e+33  \\
 53 &  J1813-1246   &  GRS                         &  4.34e+04 & 9.3e+11  & 6.2e+36  \\
 54 &  J0631+1036   &  GRS(F)                      &  4.36e+04 & 5.55e+12 & 1.7e+35  \\
 55 &  J0940-5428   &  *                           &  4.22e+04 & 1.72e+12 & 1.9e+36  \\
 56 &  J0631+1036   &  GRS(F)                      &  4.36e+04 & 5.55e+12 & 1.7e+35  \\
 57 &  J1524-5706   &  *                           &  4.96e+04 & 2.02e+13 & 1.0e+34  \\
 58 &  J1412-6145   &  *                           &  5.06e+04 & 5.64e+12 & 1.2e+35  \\
 59 &  J1809-1917   &  XRS (PWN)                   &  5.13e+04 & 1.47e+12 & 1.8e+36  \\
 60 &  J1737-3137   &  *                           &  5.14e+04 & 8e+12    & 6.0e+34  \\
 61 &  J1838-0453   &  *                           &  5.22e+04 & 6.72e+12 & 8.3e+34  \\
 62 &  J1055-6028   &  GRS                         &  5.35e+04 & 1.74e+12 & 1.2e+36  \\
 63 &  J1702-4128   &  *                           &  5.51e+04 & 3.12e+12 & 3.4e+35  \\
 64 &  J1422-6138   &  *                           &  5.58e+04 & 5.81e+12 & 9.6e+34  \\
 65 &  J1841-0345   &  *                           &  5.59e+04 & 3.48e+12 & 2.7e+35  \\
 67 &  J0633+0632   &  GRS(F)                      &  5.92e+04 & 4.92e+12 & 1.2e+35  \\
 68 &  J1429-5911   &  GRS(F)                      &  6.02e+04 & 1.9e+12  & 7.7e+35  \\
 69 &  J1406-6121   &  *                           &  6.17e+04 & 3.45e+12 & 2.2e+35  \\
 70 &  J1938+2213   &  *                           &  6.2e+04  & 2.69e+12 & 3.7e+35  \\
 71 &  J0248+6021   &  GRS(F)                      &  6.24e+04 & 3.5e+12  & 2.1e+35  \\
 72 &  J1541-5535   &  *                           &  6.25e+04 & 4.77e+12 & 1.1e+35  \\
 73 &  J1413-6205   &  GRS(F)                      &  6.28e+04 & 1.76e+12 & 8.3e+35  \\
 74 &  J1806-2125   &  *                           &  6.29e+04 & 7.74e+12 & 4.3e+34  \\
 75 &  J1105-6107   &  XRS                         &  6.33e+04 & 1.01e+12 & 2.5e+36  \\
 76 &  J1459-6053   &  GRS(F)                      &  6.47e+04 & 1.63e+12 & 9.1e+35  \\
 77 &  J1850-0026   &  *                           &  6.75e+04 & 2.58e+12 & 3.3e+35  \\
 78 &  J0534-6703   &  EXGAL:LMC                   &  6.78e+04 & 2.81e+13 & 2.8e+33  \\
 79 &  J0146+6145   &  XRS                         &  6.91e+04 & 1.33e+14 & 1.2e+32  \\
 80 &  J1954+2836   &  GRS(F)                      &  6.94e+04 & 1.42e+12 & 1.0e+36  \\
 81 &  J1636-4440   &  *                           &  7.01e+04 & 3.14e+12 & 2.1e+35  \\
 82 &  J1857+0143   &  *                           &  7.1e+04  & 2.11e+12 & 4.5e+35  \\
 83 &  J1601-5335   &  *                           &  7.33e+04 & 4.29e+12 & 1.0e+35  \\
 84 &  J1828-1101   &  *                           &  7.71e+04 & 1.05e+12 & 1.6e+36  \\
 85 &  J0901-4624   &  *                           &  8e+04    & 6.29e+12 & 4.0e+34  \\
 86 &  B1727-47     &  *                           &  8.04e+04 & 1.18e+13 & 1.1e+34  \\
 87 &  J1855+0527   &  *                           &  8.26e+04 & 1.95e+13 & 3.9e+33  \\
 88 &  J1928+1746   &  *                           &  8.26e+04 & 9.64e+11 & 1.6e+36  \\
 89 &  J1847-0130   &  *                           &  8.33e+04 & 9.36e+13 & 1.7e+32  \\
 90 &  J1705-3950   &  *                           &  8.34e+04 & 4.45e+12 & 7.4e+34  \\
 91 &  J1814-1744   &  *                           &  8.46e+04 & 5.51e+13 & 4.7e+32  \\
 92 &  J1738-2955   &  *                           &  8.58e+04 & 6.1e+12  & 3.7e+34  \\
 93 &  J1638-4608   &  *                           &  8.56e+04 & 3.83e+12 & 9.4e+34  \\
 94 &  J1803-2149   &  GRS(F)                      &  8.63e+04 & 1.46e+12 & 6.4e+35  \\
 95 &  B1916+14     &  *                           &  8.81e+04 & 1.6e+13  & 5.1e+33  \\
 96 &  B0611+22     &  *                           &  8.93e+04 & 4.52e+12 & 6.2e+34  \\
 97 &  J1718-3825   &  GRS(F),GRS(H)               &  8.95e+04 & 1.01e+12 & 1.3e+36  \\
 98 &  J1028-5819   &  GRS(F)                      &  9e+04    & 1.23e+12 & 8.3e+35  \\
 99 &  J1913+0446   &  *                           &  9.18e+04 & 2.15e+13 & 2.6e+33  \\
100 &  J1558-5756   &  *                           &  9.54e+04 & 1.46e+13 & 5.2e+33  \\
101 &  J1531-5610   &  *                           &  9.71e+04 & 1.09e+12 & 9.1e+35  \\
102 &  J1909+0912   &  *                           &  9.87e+04 & 2.86e+12 & 1.3e+35  \\
103 &  J2216+5759   &  *                           &  9.62e+04 & 5.44e+12 & 3.7e+34  \\
\hline
\enddata
\tablenotetext{\ast}{List of all pulsars with spin  down ages $< 10^5$~year, with no
  associated supernova remnant. The list is in ascending order of age,
  and the  surface magnetic  field (${\bf B}_{surf}$),  and rotational
  energy  ($\edot_{\rm rot}$)  of  the   pulsar  are  provided.  Extra-galactic
  sources, those associated with  pulsar wind nebulae (PWN), magnetars
  (SGRs or AXPs) or any other spatially coincident X-ray source (XRS),
  or Gamma-ray  source (GRS) discovered  by Fermi  (F) or Hess  (H) is
  indicated. For  further details  of pulsar  properties, associations
  and references, please see the ATNF pulsar catalogue.}
\label{tab01}
\end{deluxetable}
\begin{deluxetable}{llllll}
\tabletypesize{\footnotesize}
\rotate
\tablecolumns{6}
\tablewidth{0pt}
\tablecaption{Pulsar studies with high sensitivity {\bf \em Radio Telescopes}}
\tablehead{
           \colhead{}                                           &
           \colhead{Name \& Location}                           &
           \colhead{Type}                                       &
           \colhead{Operating Frequency\tablenotemark{\dagger}} &
           \colhead{Pulsar Studies\tablenotemark{\ast}}         &
           \colhead{Remarks}                                    
          }
\startdata
1  & Arecibo Telescope, Puerto Rico     & Single dish             &   0.3--10 GHz      & 1, 2, 3  &  The world's largest single-dish; 305~m\\
2  & Parkes Radio Telescope, Australia  & Single dish             &   0.7--26 GHz      & 1, 2, 3  &  64~m dish\\
3  & GBT, Green Bank, USA               & Single dish             &   0.29--115.3 GHz  & 2, 1, 3  &  World's largest (100~m) fully\\
   &                                    &                         &                    &          &  steerable single-dish\\
4  & Effelsberg Telescope, Germany      & Single dish             &   0.4--95 GHz      & 2, 1, 3  &  100~m dish\\
5  & Lovell Telescope, England          & Single dish             &   0.4--6 GHz       & 1, 2, 3  &  76~m dish\\
6  & GMRT, Pune, India                  & Interferometer          &   150--1420 MHz    & 3, 1, 2  &  30 dishes (45~m); largest telescope at\\
   &                                    &                         &                    &          &  meter wavelengths, SKA pathfinder\\
7  & Nancay Radio Telescope, France     & Kraus-type design       &   1.1--3.5 GHz     & 1, 2     &  single-dish antenna equivalent to that of\\
   &                                    &                         &                    &          &  a 94-m-diameter parabolic dish\\
8  & LOFAR, (mainly) Netherlands        & Dipole array            &   10--240 MHz      & 3, 1, 2  &  Low frequency array of crossed-dipole antennas\\
   &                                    &                         &                    &          &  at 1.25 to 30m wavelengths   \\
9  & Ooty radio telescope, India        & Cylindrical Paraboloid  &   326.5 MHz        & 3, 2     &  530~m $\times$ 30~m paraboloid\\
10 & Gauribidanur telescope, India      & Dipole Array            &   34 MHz           & 3, 1     &  640 dipoles in East-West direction (originally\\
   &                                    &                         &                    &          &     1000 dipoles in in a "T" configuration)\\
11 & LWA, Socorro, USA                  & Dipole array            &   10--88 MHz       & 3, 1     &  256 crossed-dipoles\\
12 & UTR--2, Ukraine                    & T-shaped dipole array   &   8--40 MHz        & 3        &  Largest telescope at decametre wavelengths\\
   &                                    &                         &                    &          &     (collecting area 150,000~m$^2$)\\
13 & MWA, Australia                     & Interferometric         &   80--300 MHz      & 3, 2, 1  &  128$\times$ 16-element cross-polar antennas\\
14 & BSA, Pushchino, Russia             & dipole array            &   100 MHz          & 3        &  16384 dipoles\\
15 & DKR-1000, Pushchino, Russia        & parabolic cylinder      &   30--120 MHz      & 3        &  1000~m $\times$ 40~m cylindrical\\
   &                                    &                         &                    &          &    paraboloids (East-West and North-South) \\
16 & HartRAO, South Africa              & Single dish             &   1.66--23 GHz     & 2        &  26~m dish\\
17 & WSRT, Netherlands                  & Interferometer          &   0.3--8.5 GHz     & 2        &  14 dish $\times$ 25~ms\\
%
%
\enddata
\tablenotetext{\ast}{Pulsar  studies  could  be  generally  linked  to
  primarily  three  observational kinds  ---  1:  Pulsar searches;  2:
  Pulsar  timing;  and 3:  Study  of  pulsar  emission mechanisms  and
  interstellar medium properties.  The order as well as the main types
  of  studies that  have been  (or potentially  could be)  carried out
  using  a  particular  telescope,  are  subjective,  and  entirely  a
  reflection of the author's perception.}
\tablenotetext{\dagger}{The operating frequency  range is nominal, and
  the observing  frequency range may  vary depending on  the receivers
  and back-end.}
\label{tab02}
\end{deluxetable}
\begin{deluxetable}{lllll}
\tabletypesize{\footnotesize}
\rotate
\tablecolumns{5}
\tablewidth{0pt}
\tablecaption{Neutron Star studies with {\bf \em High-Energy Instruments}\tablenotemark{\ast}}
\tablehead{
           \colhead{}               &
           \colhead{Mission}        &
           \colhead{Launch Date}    &
           \colhead{Status}         &
           \colhead{Neutron Star Studies}
          }
\startdata
 1 & Astrosat & current &           & cyclotron line, KHz QPOs with large area and broad energy coverage \\
 2 & Chandra  & current &           & accurate position/identification, faint source flux measurement, high resolution spectroscopy \\
 3 & Fermi    & current &           & frequency evolution, accretion torque, magnetar outbursts \\
 4 & Integral & current &           & HMXB outbursts, Cyclotron line evolution, Highly absorbed systems \\
 5 & MAXI     & current &           & accreting NS systems, orbital coverage for spectral study of bright sources \\
 6 & NuStar   & current &           & cyclotron line - new sources, spectral measurement \\
 7 & Suzaku   & current &           & cyclotron line, broad band spectroscopy \\
 8 & Swift    & current &           & super-giant Fast X-ray Transients, all sky monitoring in hard X-rays \\
 9 & XMM      & current &           & iron line shape, high resolution and high throughput spectroscopy, M/R from line during bursts \\
10 & NICER    & 2017    & approved  & EOS, soft X-ray spectral and timing studies with large area \\
11 & HXMT     & 2017    & approved  & \\
12 & eXTP     & 2025    & potential & soft X-ray imaging, polarimetry, perhaps EOS \& strong gravity \\
13 & Athena   & 2028    & approved  & \\
14 & eRosita  &         & approved  & medium energy X-ray survey, new members/population of sources? More magnetars? \\
15 & POLIX    &         & approved  & magnetic field structures, emission mechanism in accreting and young NS. \\
\enddata
\tablenotetext{\ast}{Authentic list of current and past observatories are at
{\tt http://heasarc.gsfc.nasa.gov/docs/observatories.html}}
\label{tab03}
\end{deluxetable}

\end{document}